\documentclass[twocolumn]{aastex631}

\usepackage{soul}

\usepackage{CJK}
\usepackage{amsmath} 
\usepackage{amsbsy}

\begin{document}
\begin{CJK*}{UTF8}{gbsn}

\title{The Structure and Morphology of Galaxies during the Epoch of Reionization Revealed by JWST}

\author[0000-0003-3995-4859]{Wen Sun (孙文)}
\affiliation{Kavli Institute for Astronomy and Astrophysics, Peking University, Beijing 100871, China}
\affiliation{Department of Astronomy, School of Physics, Peking University, Beijing 100871, China}

\author[0000-0001-6947-5846]{Luis C. Ho}
\affiliation{Kavli Institute for Astronomy and Astrophysics, Peking University, Beijing 100871, China}
\affiliation{Department of Astronomy, School of Physics, Peking University, Beijing 100871, China}

\author[0000-0001-5105-2837]{Ming-Yang Zhuang (庄明阳)}
\affiliation{Department of Astronomy, University of Illinois Urbana-Champaign, Urbana, IL 61801, USA}

\author{Chao Ma (马超)}
\affiliation{Kavli Institute for Astronomy and Astrophysics, Peking University, Beijing 100871, China}
\affiliation{Department of Astronomy, School of Physics, Peking University, Beijing 100871, China}

\author{Changhao Chen (陈昌灏)}
\affiliation{Kavli Institute for Astronomy and Astrophysics, Peking University, Beijing 100871, China}
\affiliation{Department of Astronomy, School of Physics, Peking University, Beijing 100871, China}

\author[0000-0001-8496-4162]{Ruancun Li (李阮存)}
\affiliation{Kavli Institute for Astronomy and Astrophysics, Peking University, Beijing 100871, China}
\affiliation{Department of Astronomy, School of Physics, Peking University, Beijing 100871, China}

\email{sunwen@stu.pku.edu.cn}

\begin{abstract}
We analyze 347 galaxies at redshift $4<z<9.5$ using JWST observations from the CEERS program by fitting a two-dimensional parametric model simultaneously to the seven-filter NIRCam images to measure the overall structural parameters and quantify the global properties of the galaxies in the rest-frame optical band. Particular attention is devoted to deriving robust uncertainties that include, among other factors, the influence of cosmological surface brightness dimming and resolution effects. Using the global S\'ersic index ($n < 1.5$) and observed axial ratio ($q < 0.6$) as a guide, we place a conservative lower limit of $\sim 45\%$ on the incidence of galactic disks. Galaxies follow a relation between rest-frame optical luminosity and effective radius in the redshift range $4<z<9.5$, as well as separately over the intervals $4 < z < 5$ and $5 \leq z < 9.5$, with a very similar slope but a marginally lower zero point in the higher redshift bin ($R_e = 0.69 \pm 0.05$~kpc) compared to the lower redshift bin ($R_e = 0.91 \pm 0.04$~kpc). Within the limitations of the current sample size, we find no significant redshift evolution of $n$ or $R_e$ at these early epochs.
\end{abstract}


\section{Introduction} \label{sec:intro}

The structure and morphology of galaxies encode important clues about their formation mechanism and evolutionary history. After their initial formation, the continued accumulation of gas fuels star formation in the outer regions of galaxies \citep[e.g.,][]{2021ApJ...919..139W}. Subsequent mergers and tidal interactions facilitate the transformation of galaxy morphology, size, and concentration \citep[e.g.,][]{2001ApJ...553...90V, 2009ApJ...699L.178N, 2010A&A...509A..78D, 2012ApJ...747...34B, 2013MNRAS.428.1088M, 2014ARA&A..52..291C, 2016ARA&A..54..761S}. Multiple pathways can lead to the formation of galactic bulges, including the early, rapid collapse of centrally concentrated gas, inward migration of large, star-forming clumps initially born in an extended disk, and galaxy-galaxy mergers (e.g., \citealt{2003ApJS..147....1C, 2014ApJ...780...57B, 2014ARA&A..52..291C}).
 
Studies of the fraction of galaxies of various morphological types show that the Hubble sequence was not yet fully established at $z \gtrsim 2$: disk galaxies and ellipticals become as common as peculiar galaxies only at $z \lesssim 1.5$ \citep[e.g.,][]{1996ApJS..107....1A, 2005ApJ...620..564C, 2008MNRAS.386..909C, 2009MNRAS.397..208C, 2021ApJ...919..139W}. Using the \cite{1968adga.book.....S} function to parameterize the global radial light distribution, an approximate proxy to describe galaxy morphology, it is apparent that star-forming galaxies at $z \approx 4-6$ are late-type systems with typical S\'ersic indices $n \approx 1-1.5$ \citep{2015ApJS..219...15S}. S\'ersic $n$ increases systematically toward lower redshifts, such that by $z \approx 1$ early-type galaxies, characterized by large $n$, become the predominant morphological class of massive galaxies \citep[e.g.,][]{2013MNRAS.428.1460B}.

Galaxy size grows over time through major and minor mergers, inside-out star formation, and gas accretion. Change in size with redshift is now well characterized at $z \lesssim 4$: distant galaxies are more compact than local galaxies of the same mass or luminosity \citep[e.g.,][]{2005ApJ...626..680D, 2007MNRAS.382..109T, 2008ApJ...687L..61B, 2013ApJ...775..106C, 2019ApJ...887..113W}. It has been argued that galaxy growth, especially for the most massive members of the population, principally involved dry, minor mergers \citep[e.g.,][]{2009ApJ...699L.178N, 2012ApJ...747...34B, 2017MNRAS.465..722F}, although some contend that the role of major mergers cannot be overlooked (e.g., \citealt{2017ApJ...836...75D}). Observation of the rest-frame ultraviolet (UV) emission of galaxies at $z > 4$ suggests that at fixed luminosity or mass, galaxy sizes slowly decrease toward higher redshift \citep[e.g.,][]{2004ApJ...611L...1B, 2004ApJ...600L.107F, 2010ApJ...709L..21O, 2015ApJS..219...15S}. These results resonate with expectations from the scenario proposed by \citet{1980MNRAS.193..189F}, which posits that galactic disks form within dark matter halos that acquire angular momentum from tidal torques. Baryons initially share the same specific angular momentum as the dark matter, and angular momentum is conserved as the baryons collapse and cool to form a disk. In this scenario, the expected size scales as $R_e \propto H(z)^{-1}$ at a fixed halo circular velocity or $R_e \propto H(z)^{-\frac{2}{3}}$ at a fixed halo mass \citep{1980MNRAS.193..189F, 1998MNRAS.295..319M, 2004ApJ...600L.107F}, where $H(z)$ is the Hubble parameter at redshift $z$. 

Whether and how galaxy size evolves at high redshift remain controversial. For instance, while \citet{2019ApJ...887..113W} show that mergers contribute to the increase in galaxy size from $z = 7$ to $z = 1$, \cite{2016MNRAS.457..440C} find little evidence that galaxy size in the rest-frame UV evolves at high redshift. Most studies of galaxy structure evolution have been limited to redshifts less than $\sim 4$ \citep[e.g.,][]{1995ApJ...453...48D, 1996ApJS..107....1A, 2004AJ....127..131S, 2005ApJ...631..101P, 2008MNRAS.386..909C, 2011ApJ...742...96W, 2013ApJ...774...47L, 2013MNRAS.433.1185M, 2014ApJ...788...28V, 2015Sci...348..314T, 2019MNRAS.484.5170Z, 2021ApJ...919..139W, 2022ApJ...929..121C}. At higher redshifts, the detailed properties of galaxies are challenging to resolve because the galaxies are faint, small, and impacted by surface brightness dimming. Equally seriously, the reddest band accessible with the Hubble Space Telescope (HST), $\sim 1.6\,\mu$m, can no longer capture the rest-frame optical emission of galaxies that better traces their overall stellar population. Thus, although deep near-infrared HST imaging has greatly advanced our understanding of the rest-frame UV properties of galaxies at $z > 4$ (e.g., \citealt{2010ApJ...709L..16O, 2011ApJS..197...35G, 2011ApJ...727L..39T, 2017ApJ...837...97L}), we still critically lack robust measurements of the morphologies and structural parameters of galaxies during their formative years.

The James Webb Space Telescope (JWST: \citealt{2023arXiv230101779M, 2023PASP..135d8001R}) has ushered in a new era for probing many aspects of the early epoch of galaxy evolution, including their morphology and structure \citep[e.g.,][]{2022ApJ...938L...2F, 2023ApJ...946L..15K, 2023arXiv230207234T}. With a 6.5~m primary mirror, JWST has 7 times the light-gathering power of HST. Equally importantly, JWST significantly extends the long-wavelength sensitivity of HST while maintaining excellent image quality, a crucial combination for resolving the internal stellar properties of high-redshift galaxies. For example, the reddest wide filter (F444W) of the Near Infrared Camera \citep[NIRCam;][]{2023PASP..135b8001R} enables the study of the rest-frame optical light of galaxies up to $z \approx 9.5$. Contrary to a preconception established by previous HST studies, recent results from JWST indicate that regular disks are common in galaxies at $z \gtrsim 2$ \citep[e.g.,][]{2022ApJ...938L...2F, 2022arXiv220806516J, 2022arXiv220801630N, 2023ApJ...942L..42R}, suggestive of an early emergence of the Hubble sequence, although the fraction of purely irregular galaxies, showing no signs of a disk or spheroidal component, rises toward higher redshift \citep{2023ApJ...946L..15K}. These studies also discuss a variety of systematic trends involving the S\'ersic index, with no clear consensus yet emerging from the initial explorations of JWST data.

In this study, we investigate the structural parameters of a sample of galaxies at $z = 4-9.5$ selected from the Cosmic Evolution Early Release Science (CEERS; \citealt{2023ApJ...946L..13F}) survey, which covers $\sim 100$~arcmin$^2$ of the Extended Groth Strip \citep[EGS; ][]{Davis_et_al.(2007)} with JWST imaging and spectroscopy using NIRCam. We use the sample, which traverses into the epoch of reionization, to investigate the incidence of galactic disks, the luminosity-size relation of galaxies, and the possible evolution of size and morphology with redshift. Although the CEERS data already have been the target of recent studies with broadly similar science goals, as mentioned above, our approach differs in several key aspects. Despite the impressive capabilities of JWST, securing accurate structural parameters with robust uncertainties for high-redshift galaxies is still a non-trivial challenge. Some high-redshift sources remain barely resolved even with JWST. We examine the potential impact of the choice of image pixel scale used in the data reduction process, and we systematically investigate different treatments of the point-spread function (PSF) to arrive at an optimal solution. As the appearance of a galaxy changes with redshift as a result of cosmological surface brightness dimming and image resolution \citep[e.g.,][]{1996AJ....112..369G, 1997AJ....114.1741H, 2003ApJS..147....1C, 2008ApJS..175..105B, 2013MNRAS.435..623V, 2016ApJ...824..112D}, we design realistic mock experiments to quantify the contribution of these effects to the final error budget. Lastly, galaxy morphology and structure depend on wavelength, as a consequence of internal variations in stellar population and dust attenuation \citep{2002ApJS..143..113W, 2007ApJ...659..162T, 2012MNRAS.421.1007K, 2013MNRAS.430..330H, 2013MNRAS.435..623V}. It is important to take this into consideration when measuring the structural parameters of galaxies across different bands, as is the case for the seven-band images of CEERS. We adopt {\tt GALFITM} \citep{2013MNRAS.430..330H, 2013MNRAS.435..623V} to construct a wavelength-dependent model to fit the multiband images simultaneously, an approach shown to deliver more accurate photometric and structural parameters, especially in the regime of low signal-to-noise ratio \citep{2022A&A...664A..92H}. This technique has been put into practice in a variety of applications (e.g., \citealt{2022ApJ...934..130Z}), including CEERS itself \citep{2023ApJ...946L..15K}.

The paper is organized as follows. Section~\ref{sec:obser} describes the observations used in this work, the procedure for data reduction, the strategy to construct the PSF, and sample selection. Section~3 introduces our method of image analysis, measurement results, mock simulations to quantify redshift effects, and error analysis. Section~4 discusses the main implications for the incidence of galactic disks, the luminosity-size relation, and the possible evolution of the galaxy structural parameters. A summary appears in Section~\ref{sec:conclusions}. We assume the latest Planck flat $\Lambda$CDM cosmology with $H_0 = 67.36$~km~s$^{-1}$~Mpc$^{-1}$, $\Omega_m = 0.3153$, and $\Omega_{\Lambda} = 0.6847$ \citep{2020A&A...641A...6P}. All magnitudes are in the absolute bolometric (AB) system \citep[][]{1983ApJ...266..713O}.

\section{Observations} \label{sec:obser}

\subsection{CEERS Data} \label{subsec:CEERSData}

We analyze the first four of the 10 CEERS NIRCam pointings (CEERS1, CEERS2, CEERS3, and CEERS6) obtained on 21 June 2022, which cover $34.5$~arcmin$^2$. Each pointing uses the filters F115W, F150W, and F200W in the short-wavelength (SW) channel and the filters F277W, F356W, F410M, and F444W in the long-wavelength (LW) channel. The total exposure time per filter is typically 2835~s for pixels observed in all three dithers, except for F115W, which has 2 times longer exposure time to increase the depth of the wavelength range bluer than the Lyman-break at $z > 10$. The CEERS survey is optimized to study the abundance and physical nature of galaxies in the early Universe ($z > 10$) and the processes of galaxy assembly and black hole growth at $z \approx 1-10$.

\begin{figure*}
\begin{center}
\includegraphics[width=1.0\textwidth]{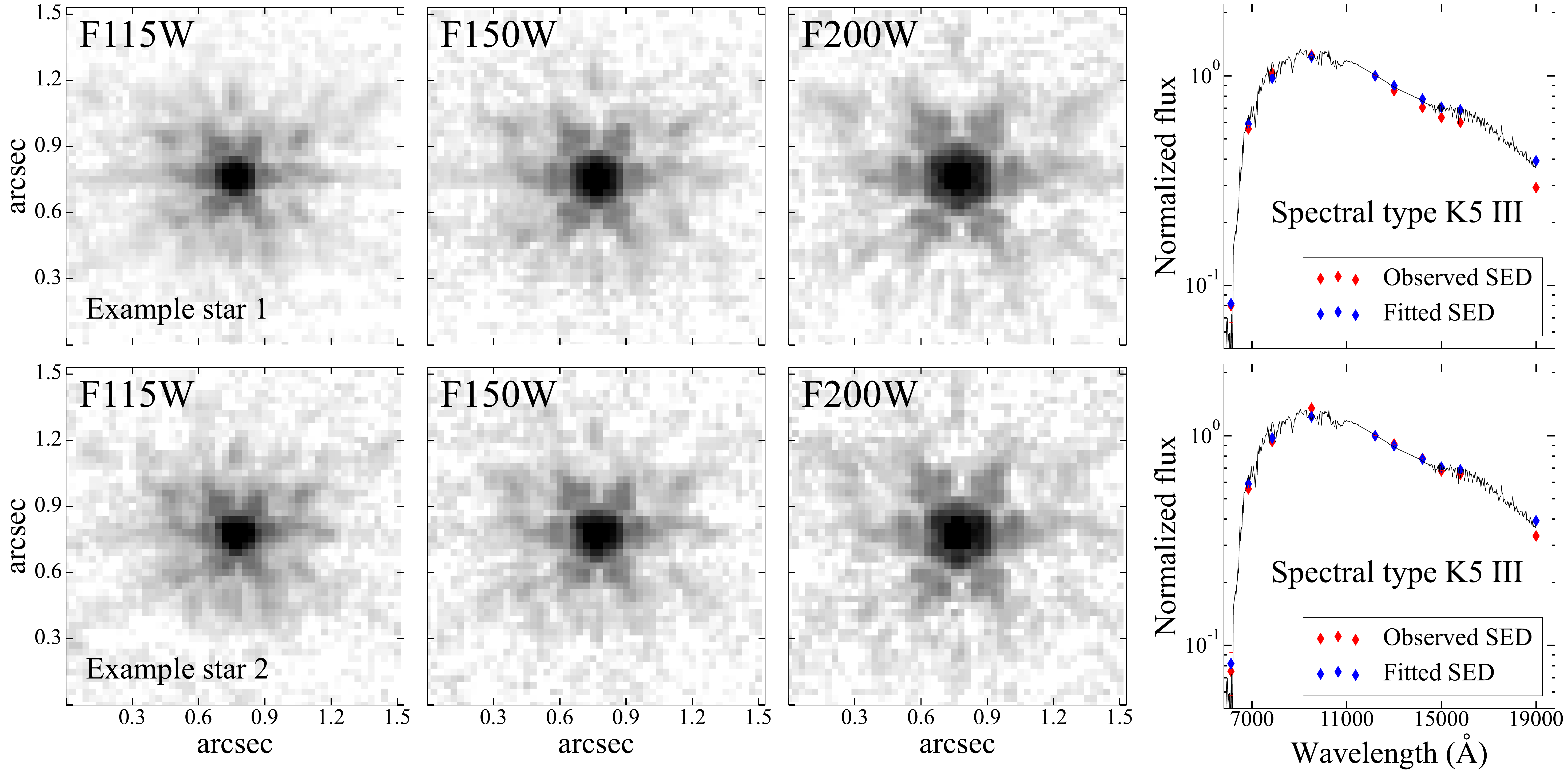}
\end{center}
\caption{Two example stars in the CEERS1 field. From left to right are their images in the F115W, F150W, and F200W bands, and their photometric data points overplotted on the best-fit stellar spectrum. \label{fig:sed_compare}}
\end{figure*}

\subsection{Data Reduction} \label{subsec:DataReduction}

\citet{2023ApJ...946L..12B} describe the data reduction of the first public release (Data Release~0.5) of the CEERS NIRCam imaging\footnote{\url{https://ceers.github.io/dr05.html}}, which includes custom procedures beyond those of the standard JWST pipeline for removal of $1/f$ noise, wisps, and snowballs from the countrate maps, astrometric calibration, image coaddition and mosaicing, and background subtraction. The final images are drizzled to a pixel scale of $0\farcs030$.

In view of the large distances and intrinsically compact sizes of high-redshift galaxies \citep[e.g.,][]{Damjanov_et_al.(2009), Cimatti_et_al.(2012)} we place special emphasis on achieving the highest possible resolution in order to obtain the most reliable measurements of source structure and morphology. The possibility that some galaxies may host an active galactic nucleus (e.g., \citealt{2023arXiv230311946H}) further underscores the need to resolve accurately their internal substructure. While the default pixel size of $0\farcs030$ provided by CEERS Data Release~0.5 Nyquist samples the PSF of the bands in the LW channel, which have full-width at half-maximum (FWHM) $\sim 0\farcs119-0\farcs160$, it undersamples the three SW channel bands, which have a PSF of ${\rm FWHM} = 0\farcs0605$, $0\farcs0647$, and $0\farcs0750$ for F115W, F150W, and F220W, respectively \citep{Zhuang_Shen(2023)}. To investigate the potential effects of the output pixel scale after drizzling on our measured results, we perform a custom set of data reduction to generate images with a finer output pixel scale for the bands in the SW channel. Our calibration procedures are similar to those of \citet{2023ApJ...946L..12B}. Starting with the Stage~2 data products acquired from the STScI MAST Portal\footnote{\url{https://archive.stsci.edu/}}, we process them using version~1.7.2 of the JWST Calibration Pipeline with the mapping file jwst\_0965.pmap. Prior to removing $1/f$ noise, we use {\tt Photutils} to mask sources in four iterations, dilating them in between iterations by 45, 35, 29, and 25 pixels. We adopt larger dilation sizes than \citet{2022arXiv220712474F} to better mask extended objects. In the resampling step of the Stage~3 pipeline, we drizzle the individual images to $0\farcs015\,{\rm pixel}^{-1}$ for the SW channel and $0\farcs030\,{\rm pixel}^{-1}$ for the LW channel. As our image analysis method (Section~4.1) requires an identical pixel scale across all filters, prior to model fitting we reproject the mosaics in the three SW filters to the same final pixel scale of the images of the LW filters ($0\farcs030\,{\rm pixel}^{-1}$). We use the Gaia DR3 source catalog\footnote{\url{https://gea.esac.esa.int/archive/}} \citep{2016A&A...595A...1G} for astrometric calibration. After masking bad pixels and sources, we use {\tt Photutils.Background2D} to estimate and subtract any residual background. We do not remove ``snowballs'' (circular defects) from the NIRCam mosaics \citep[e.g.,][]{2022arXiv220712474F, 2022ApJ...938L..14M, 2023ApJ...946L..12B, 2023PASP..135d8001R}, but we carefully inspect all the images to confirm that our sources of interest are not contaminated by these artifacts. The finer drizzle pixel scale has a minimal impact on the source magnitude and size, but the effect on the S\'ersic index can be substantial (Section~\ref{appen:pixelscale}).

The flux calibration procedure of the JWST pipeline is still evolving. \citet{2022RNAAS...6..191B} evaluate the flux calibration of NIRCam using globular cluster data from the Resolved Stellar Populations Early Release Science program, concluding that the zero point for the F150W filter derived with their two-dimensional kernel density estimator technique is slightly better than those from \citet{gabriel_brammer_2022_7143382} and much better than others. However, a similar analysis is not available for the other NIRCam filters. We adopt the zero point for F150W from \citet{2022RNAAS...6..191B}, and for the other six filters, we use the zero points from \citet{gabriel_brammer_2022_7143382}.

\subsection{Point-spread Function} \label{subsec:psf}

Despite the excellent spatial resolution of JWST, some high-redshift galaxies remain barely resolved. Under these circumstances, the accuracy of the PSF model is of paramount importance for obtaining reliable source structural parameters. We detail our procedure for constructing the PSF for the CEERS fields and compare different PSF models in Appendix \ref{appen:psf}. For each NIRCam filter, we construct a PSF by stacking a number of isolated, unsaturated stars, which are identified as point-like sources in our source catalog (Section~\ref{subsec:catalog}) that have ${\rm FWHM} < 6$ pixels, as determined by fitting a two-dimensional Gaussian profile to an image cutout of size 7 times the \cite{1980ApJS...43..305K} radius of the source that is uncontaminated by neighboring sources. To separate stars from other compact sources, such as quasars and dwarf galaxies, we perform a least-squares fit between the observed SED of the candidate point-like source and stellar spectral templates from the ESO Library of Stellar Spectrum\footnote{\url{https://www.eso.org/sci/facilities/paranal/decommissioned/isaac/tools/lib.html}}. The observed SED comprises 10 photometric points, three (F115W, F150W, and F200W) from the SW channel in combination with seven additional bands covering $\lambda\approx 0.4-1.6~\mu$m \citep{2017ApJS..229...32S}. We exclude the four LW bands because the stellar spectral templates do not extend beyond $\sim 2~\mu$m. To avoid potential confusion with quasars, we do not consider templates of type O, B, and A stars, which, in any case, are expected to be rare because of their short main-sequence lifetimes and because they are faint in the near-infrared.

Figure~\ref{fig:sed_compare} shows two examples of our final sample of 18 stars, which are either G-type or K-type giants, identified in the four CEERS pointings. Because of the relatively small number of stars available, we do not produce a separate PSF for each pointing but instead combine the stars in all four pointings to produce a master, stacked PSF of high signal-to-noise ratio for each filter. PSF variations across different pointings should be small because they utilize the same dither pattern and were taken close in time \citep{2023ApJ...946L..13F}. We extract a $81 \times 81$ pixel cutout for each star, and all the star cutouts are 4 times oversampled to align their centers. We construct the final PSF by mean-combining the individual stars and resampling the oversampled images back to their original resolution. See Appendix~\ref{appen:psf} for details.

\subsection{Sample Definition} 
\label{subsec:catalog}

We employ custom {\tt Photutils} scripts for source detection and photometry. We use the LW channel data instead of the SW bands for source detection to avoid the complications of substructures arising from star-forming clumps, which, more prominent in the rest-frame UV, can split single galaxies into multiple sources. Stacking the mosaics of the four LW bands weighted by their inverse variance produces a detection image of a high signal-to-noise ratio. Sources are defined as contiguous regions of the segmentation map that contain five or more connected pixels with values larger than $1 \sigma$ above the background. A local peak must have at least 0.001 of the total source flux to be deblended as a separate object. We visually inspect all sources to verify that they have been properly deblended.

\begin{figure}[t]
\center{\includegraphics[width=1.0\linewidth]{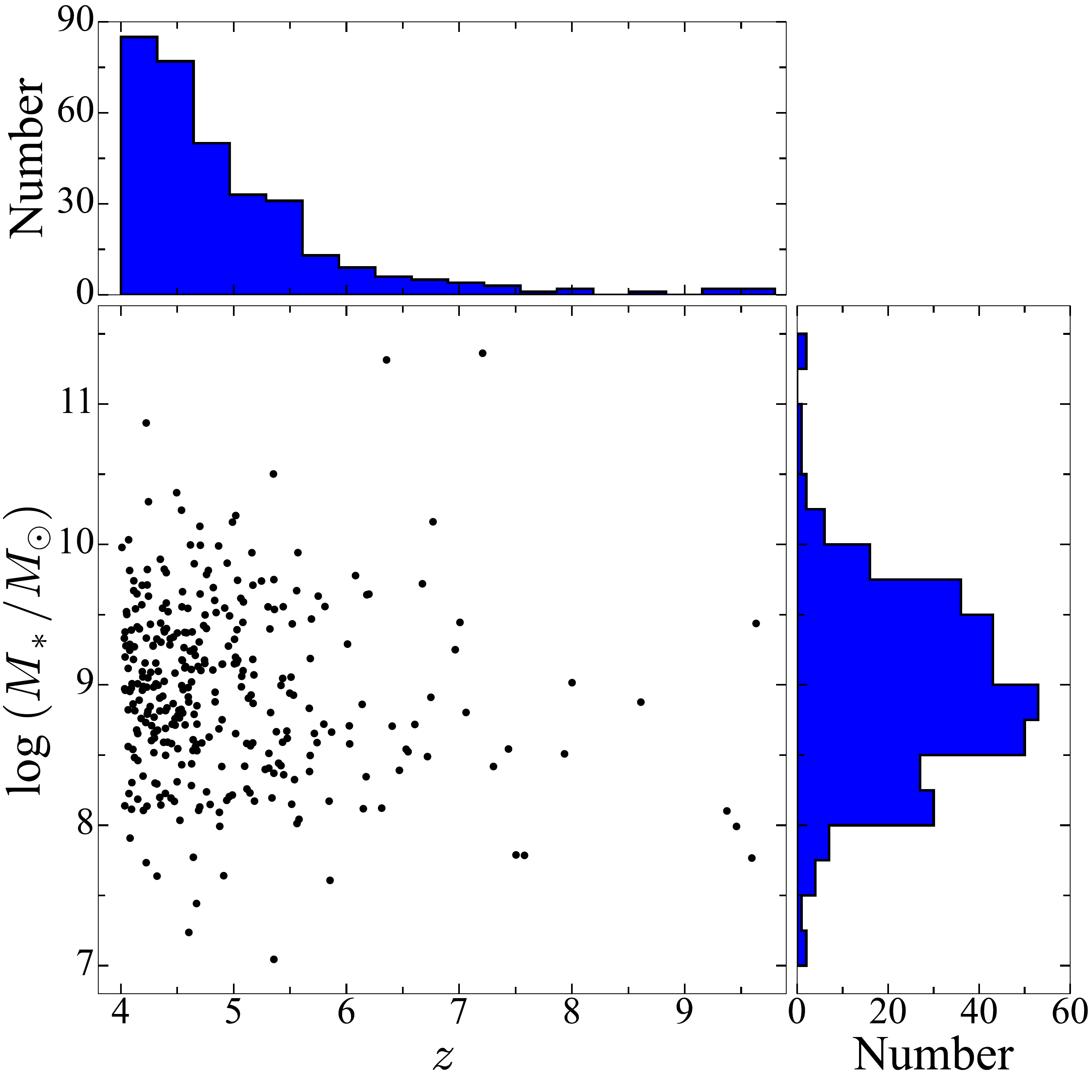}}
\caption{The distribution of redshift and stellar mass of the sample. \label{fig:mass_z_distri}}
\end{figure}

This study focuses on the subset of galaxies in CEERS with redshifts $4 < z < 9.5$ that cross-matches within $1\farcs0$ with the multiwavelength catalog of the CANDELS EGS field of \cite{2017ApJS..229...32S}, who curated 22 bands of photometry from 0.4 to $8~\mu$m, including HST WFC3 and ACS data from CANDELS. Although there are other catalogs of photometric redshifts available \citep[e.g., ][]{2019ApJ...876..110D, 2021ApJ...919..139W, 2023ApJ...942...36K}, we choose Stefanon et al.'s catalog for convenience because it provides both photometric redshifts and stellar masses. Other recent CEERS studies also follow this strategy \citep[e.g., ][]{2023ApJ...945L..10G, 2023ApJ...942L..42R}. Whenever possible, preference is given to spectroscopic redshifts. A total of 389 galaxies match our selection criteria. Since there are overlaps between different CEERS NIRCam pointings, five galaxies have multiple images; we retain the best image and discard the redundant ones. After removing galaxies that lie at the edge of the field and do not have complete imaging, and those that are located on the spikes of foreground stars or are contaminated by a nearby bright source, we are left with a final sample of 347 galaxies (Table~\ref{tab:all_par}). Figure~\ref{fig:mass_z_distri} shows the redshift and stellar mass distribution of the sample. The median uncertainty of the sample's stellar masses is $\sigma\left[\mathrm{log} \left(M_{*}/M_{\odot}\right)\right] = 0.327$ dex and that of the photometric redshifts is $\sigma(z)/(1+z) = 0.407$.

\begin{splitdeluxetable*}{cccccccccBccccccccc}
\tabletypesize{\scriptsize}
\tablewidth{0pt} 
\tablecaption{Measurements of Magnitudes and Structural Parameters \label{tab:all_par}}
\tablehead{
\colhead{Name} & \colhead{R. A.}& \colhead{Dec.} & \colhead{$z$} & \colhead{log $M_{*}$} &
\colhead{$m_{\mathrm{F115W}}$} & \colhead{$m_{\mathrm{F150W}}$} & \colhead{$m_{\mathrm{F200W}}$} & \colhead{$m_{\mathrm{F277W}}$} &
\colhead{$m_{\mathrm{F356W}}$} & \colhead{$m_{\mathrm{F410M}}$} & \colhead{$m_{\mathrm{F444W}}$}
& \colhead{Rest-frame} & \colhead{$n$} &
\colhead{$R_e$} & \colhead{$q$} &
\colhead{$\Theta$} &
\colhead{Quality} \\
\colhead{} & \colhead{($^{\circ}$)} & \colhead{($^{\circ}$)}& \colhead{}& \colhead{($M_{\odot}$)} & \colhead{(mag)} & \colhead{(mag)} & \colhead{(mag)} & \colhead{(mag)} & \colhead{(mag)} & \colhead{(mag)} & \colhead{(mag)} & \colhead{Optical Band} & \colhead{} & \colhead{(kpc)} & \colhead{} & \colhead{($^{\circ}$)} & \colhead{Flag}}
\colnumbers
\startdata
EGS$-$20901 & 215.04927 & 52.99757 & 4.87 & 8.69 $\pm$ 0.67 & 25.97 $\pm$ 0.08 & 25.86 $\pm$ 0.06 & 26.02 $\pm$ 0.05 & 25.34 $\pm$ 0.04 & 25.65 $\pm$ 0.04 & 25.24 $\pm$ 0.05 & 25.39 $\pm$ 0.06 & F356W & 1.77 $\pm$ 0.45 & 1.05 $\pm$ 0.10 & 0.33 $\pm$ 0.08 & 81.8 $\pm$ 9.7 & 1 \\
EGS$-$21112 & 215.02291 & 52.98007 & 5.32 & 10.65 $\pm$ 0.12 & 25.93 $\pm$ 0.14 & 25.51 $\pm$ 0.08 & 24.83 $\pm$ 0.03 & 24.29 $\pm$ 0.01 & 24.00 $\pm$ 0.01 & 23.65 $\pm$ 0.01 & 23.45 $\pm$ 0.01 & F356W & 1.07 $\pm$ 0.92 & 0.78 $\pm$ 0.37 & 0.85 $\pm$ 0.14 & 74.6 $\pm$ 10.0 & 2 \\
EGS$-$21844 & 215.02935 & 52.98810 & 4.55 & 9.66 $\pm$ 0.24 & 24.90 $\pm$ 0.02 & 24.70 $\pm$ 0.01 & 24.42 $\pm$ 0.01 & 24.18 $\pm$ 0.01 & 24.11 $\pm$ 0.01 & 24.22 $\pm$ 0.01 & 24.25 $\pm$ 0.01 & F356W & 1.18 $\pm$ 1.00 & 0.90 $\pm$ 0.03 & 0.74 $\pm$ 0.05 & 166.8 $\pm$ 2.2 & 1 \\
EGS$-$21879 & 215.02867 & 52.98805 & 4.43 & 9.28 $\pm$ 0.19 & 25.57 $\pm$ 0.09 & 25.67 $\pm$ 0.07 & 25.49 $\pm$ 0.04 & 25.30 $\pm$ 0.03 & 25.26 $\pm$ 0.03 & 25.38 $\pm$ 0.05 & 25.38 $\pm$ 0.07 & F356W & 1.56 $\pm$ 0.62 & 2.03 $\pm$ 0.94 & 0.17 $\pm$ 0.09 & 175.3 $\pm$ 8.9 & 1 \\
EGS$-$21944 & 214.98304 & 52.95601 & 6.47 & 10.51 $\pm$ 0.14 & 26.58 $\pm$ 0.14 & 25.97 $\pm$ 0.10 & 25.74 $\pm$ 0.07 & 25.55 $\pm$ 0.04 & 24.39 $\pm$ 0.03 & 23.86 $\pm$ 0.02 & 24.03 $\pm$ 0.02 & F444W & 6.93 $\pm$ 2.64 & 0.03 $\pm$ 0.16 & 0.53 $\pm$ 0.17 & 137.2 $\pm$ 13.5 & 2 \\
\enddata
\tablecomments{Col. (1): Object name from \citet{2017ApJS..229...32S}. Col. (2): Right ascension (J2000). Col. (3): Declination (J2000). Col. (4): Redshift. Col. (5): Stellar mass. Cols. (6)--(12): Integrated magnitudes and associated uncertainties (see Section~\ref{subsec:error}). Col. (13): Rest-frame optical band; we exclude the medium filter F410M because its wavelength coverage is less than that of the wide filters, and its central wavelength is close to that of F444W. Col. (14): S{\'e}rsic index. Col. (15): Effective radius. Col. (16): Axis ratio. Col. (17): Position angle. Col. (18): Quality flag, where 1 corresponds to $\chi^2_{\nu}\leq 1.9$ and 2 denotes $\chi^2_{\nu}> 1.9$.  (This table is available in its entirety in machine-readable format.)}
\end{splitdeluxetable*}

\begin{figure*}[ht!]
\begin{center}
\includegraphics[width=0.70\textwidth]{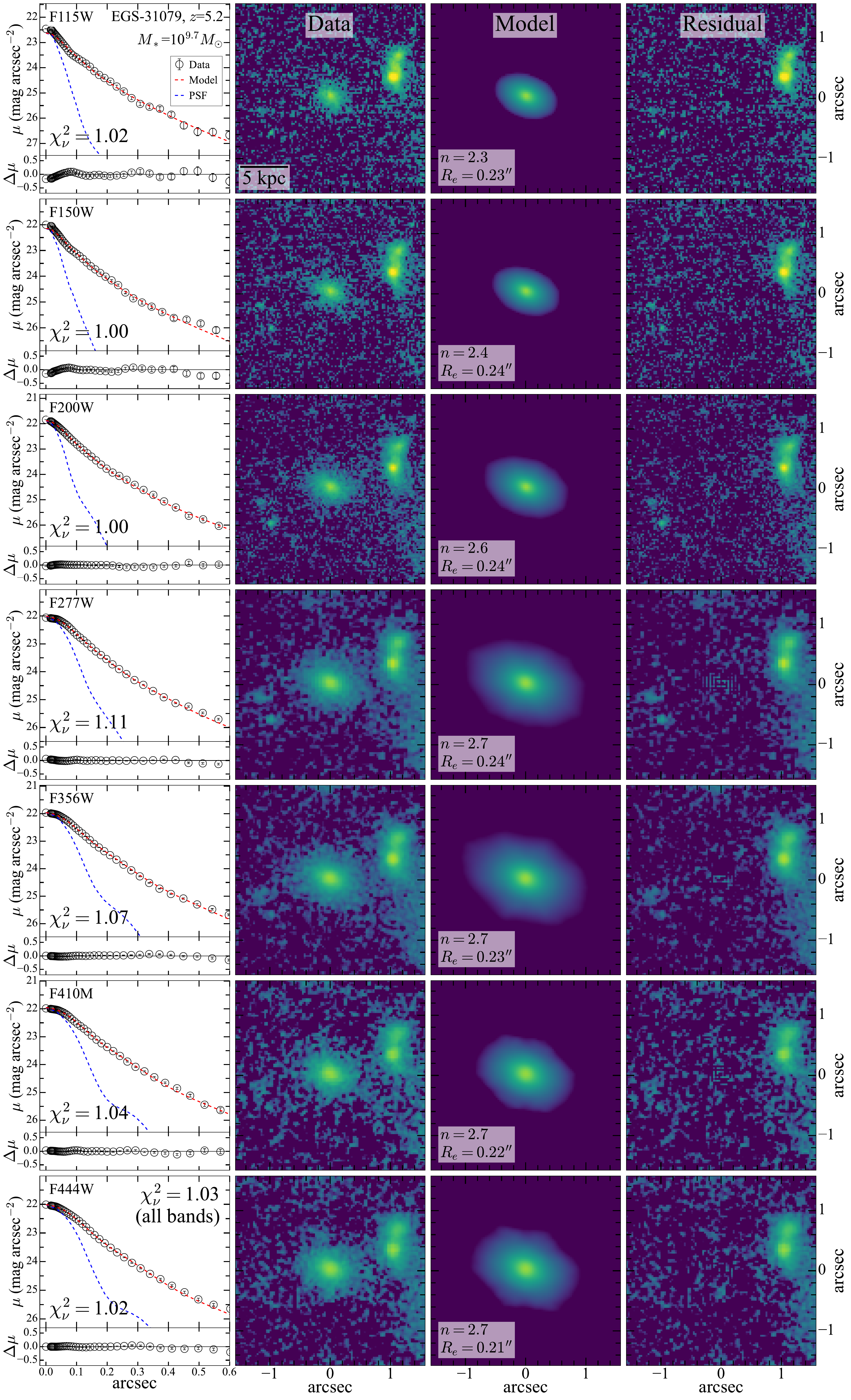}
\end{center}
\caption{Simultaneous, multiband fitting of the galaxy EGS$-$31079 \citep{2017ApJS..229...32S} at $z=5.2$. Rows from top to bottom are the results for the seven JWST/NIRCam filters F115W, F150W, F200W, F277W, F356W, F410M, and F444W. The upper panel of the left column shows the radial surface brightness profile of the galaxy (open circles with error bars), the best-fitting model (red dashed line), and the PSF model (blue dotted line). The $\chi^2_{\nu}$ from {\tt GALFITM} for each band is given in the lower-left corner, while that for all seven bands is given in the upper-right corner of the first panel in the bottom row. The lower subpanel gives the residuals between the data and the best-fit model (data $-$ model). The images show, from left to right, the original data, model, and residuals. The best-fit S{\'e}rsic index $n$ and effective radius $R_e$ are given in the third column. \label{fig:31079_ellip}}
\end{figure*}

\begin{figure*}
\begin{center}
\includegraphics[width=0.70\textwidth]{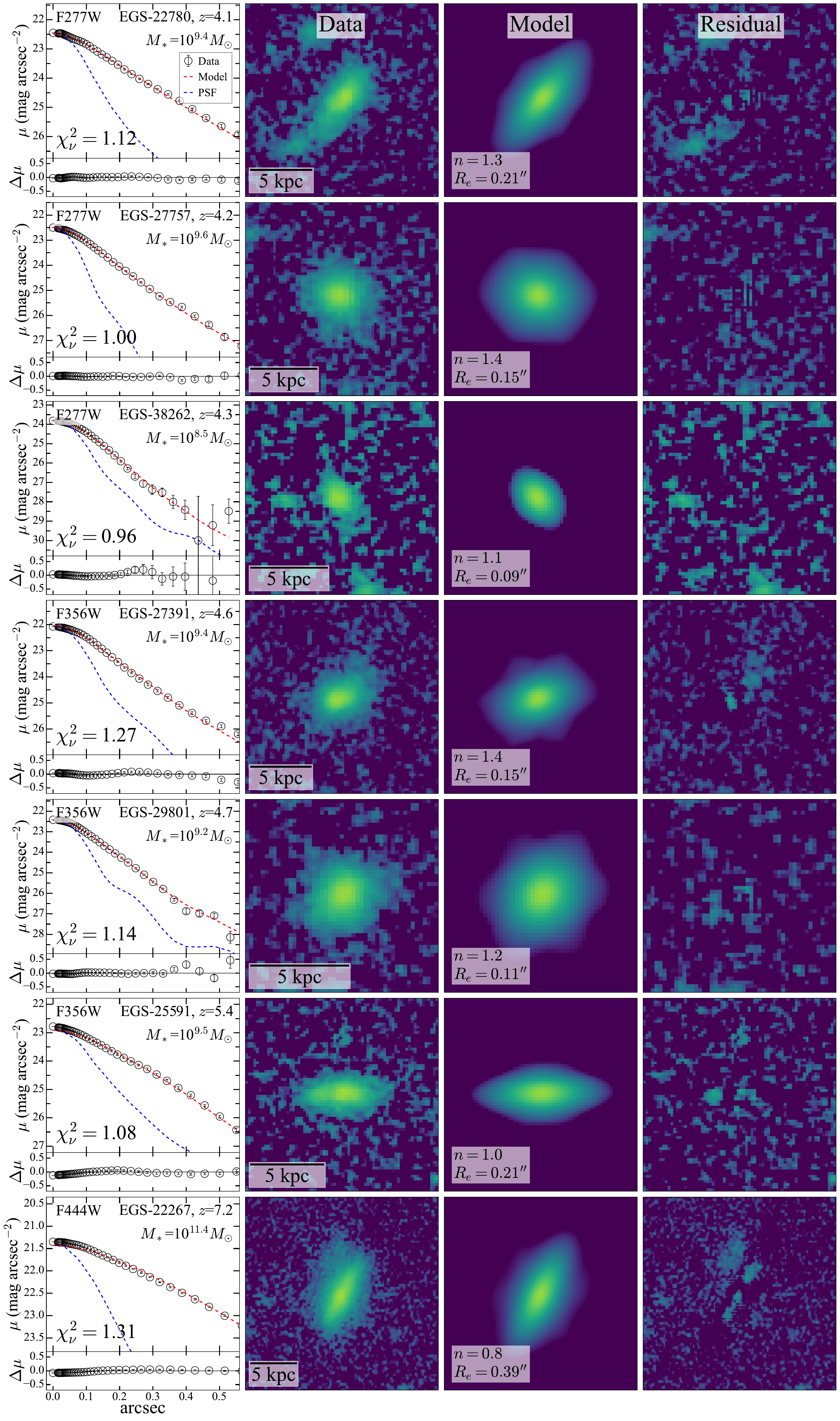}
\end{center}
\caption{Simultaneous, multiband fitting in the rest-frame optical band for example galaxies with different $M_{*}$ and relatively disky morphologies at various redshifts. The upper panel of the left column shows the radial surface brightness profile of the galaxy (open circles with error bars), the best-fitting model (red dashed line), and the PSF model (blue dotted line). The $\chi^2_{\nu}$ from {\tt GALFITM} for all seven bands is given in the lower-left corner. The lower subpanel gives the residuals between the data and the best-fit model (data $-$ model). The images show, from left to right, the original data, model, and residuals. The best-fit S{\'e}rsic index $n$ and effective radius $R_e$ are given in the third column. \label{fig:multi_ellip}}
\end{figure*}

\begin{figure*}
\begin{center}
\includegraphics[width=0.70\textwidth]{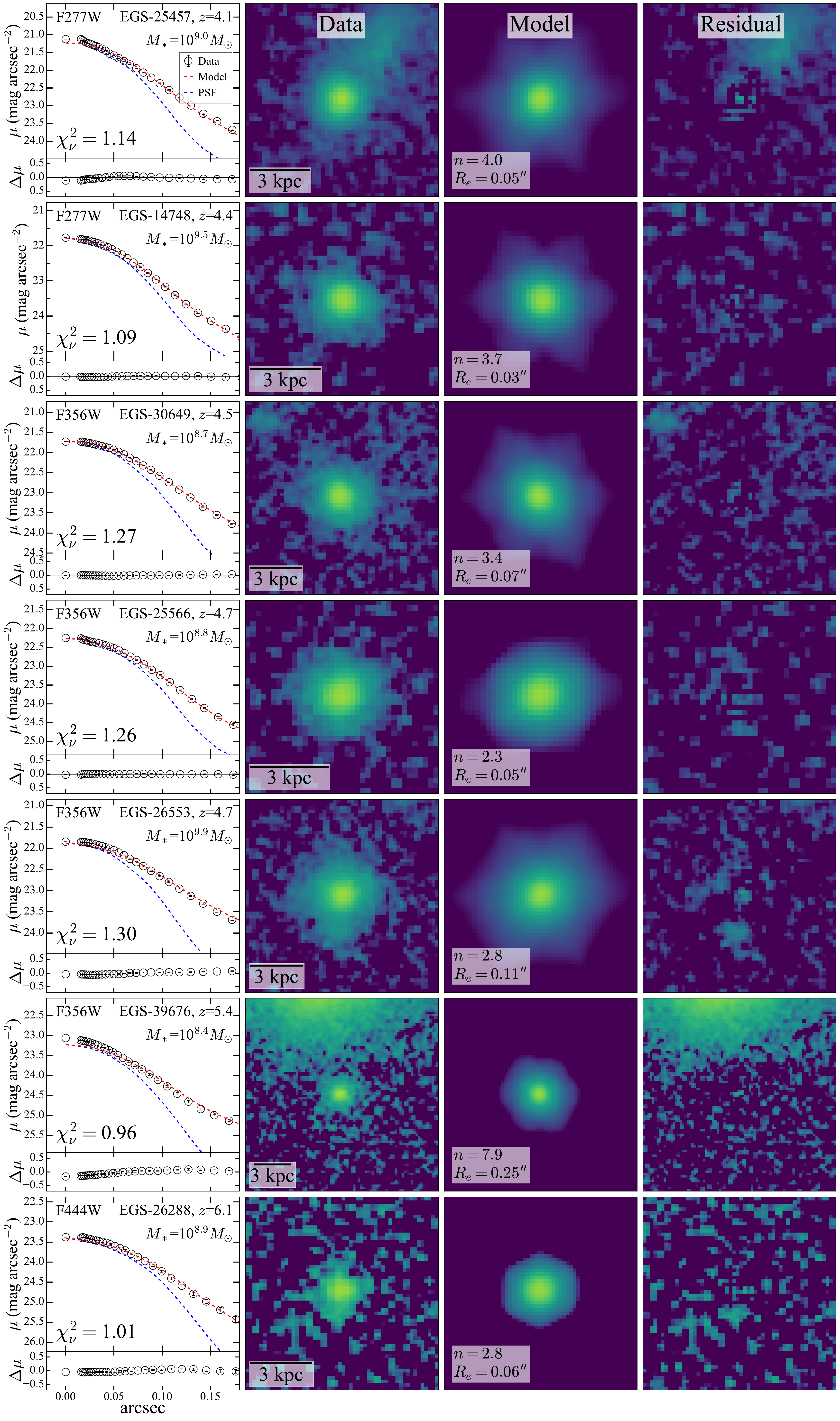}
\end{center}
\caption{As with Figure~\ref{fig:multi_ellip}, but for galaxies that are more spheroidal in morphology. The upper panel of the left column shows the radial surface brightness profile of the galaxy (open circles with error bars), the best-fitting model (red dashed line), and the PSF model (blue dotted line). The $\chi^2_{\nu}$ from {\tt GALFITM} for all seven bands is given in the lower-left corner. The lower subpanel gives the residuals between the data and the best-fit model (data $-$ model). The images show, from left to right, the original data, model, and residuals. The best-fit S{\'e}rsic index $n$ and effective radius $R_e$ are given in the third column. \label{fig:multi_ellip_1}}
\end{figure*}

\section{Image Analysis} 

\subsection{Methodology} \label{sec:method}

Owing to internal variations in stellar population and dust attenuation, galaxy morphology and structure depend on wavelength. To account for the variation of galaxy structure with wavelength, we analyze the images of all seven NIRCam bands using {\tt GALFITM} \citep{2013MNRAS.430..330H, 2013MNRAS.435..623V}, a multiband version of the widely used two-dimensional image fitting code {\tt GALFIT} \citep{2002AJ....124..266P, 2010AJ....139.2097P}. The program fits the pixel-registered images of multiple filters to produce a consistent, wavelength-dependent model of the galaxy, with the aid of a user-specified polynomial function to constrain the wavelength dependence of the structural parameters of the model components\footnote{\url{https://www.nottingham.ac.uk/astronomy/megamorph/}}. The free parameters of the model are fitted to the multiband data simultaneously by minimizing a single likelihood function. \citet{2022A&A...664A..92H} show, through a series of simulations, that this approach significantly reduces deviations from true parameter values, allows component sizes and S{\'e}rsic indices to be measured more accurately, and constrains the band-to-band parameter variations to more physical values. {\tt GALFITM} delivers better performance especially in the regime of low signal-to-noise ratio \citep{2013MNRAS.430..330H, 2013MNRAS.435..623V, 2021A&A...647A..89N, 2022A&A...664A..92H}.

After some experimentation, we find that a cutout size of 7 times the \cite{1980ApJS...43..305K} radius of the source on the detection image (Section~\ref{subsec:catalog}) can capture the galaxy outskirts while including adequate background. The center of the galaxy needs to be aligned when creating the cutouts to correct for residual offsets of up to several pixels that exist between the images from the SW and LW channels. The segmentation map is used to mask objects excluded from the fit. As the segmentation map often misses the fainter outer regions of bright sources, we dilate the segmentation image of each object by calculating a ``growth radius'' \citep{2011ApJS..197...21H, 2011ApJS..197...22L, 2013ApJ...766...47H}. From trial and error, we find that the optimal growth radius for the CEERS images can be approximated by $R = {\rm min}[8, 0.8\sqrt{N/\pi}]$, where $N$ is the number of pixels contained in the original segmentation image of the object. After masking all sources in each cutout, the local background is estimated through sigma-clipping and then removed. We confirm that the dilated mask of nearby large objects does not influence the object of interest. Similar to \citet{2023ApJ...946L..13F} and \citet{2023ApJ...946L..15K}, we use the cutout of the error array (ERR extension) from the JWST pipeline as the input sigma image, which includes not only Poisson noise from the source but also instrumental noise. The sigma image is scaled typically by a factor $\sim 1.3-2.5$, depending on the filter, such that its background pixel values roughly equal the standard deviation of the background pixel values of the science image (SCI extension).

With the background-subtracted cutout, mask, and PSF of each band in hand, we simultaneously fit the seven bands with a two-dimensional surface brightness model represented by the S{\'e}rsic function 

\begin{equation}
\Sigma(R)=\Sigma_e\exp{\left\{-\kappa\left[\left(\frac{R}{R_e}\right)^{1/n}-1\right]\right\}},
\end{equation}

\noindent
where $R_e$ is the effective radius of the galaxy that contains half of the total flux, $\Sigma_e$ is the surface brightness at $R_e$, the S\'ersic index $n$ specifies the shape of the light profile, and $\kappa$ is related to $n$ by the incomplete gamma function, $\Gamma\left(2n\right)=2\gamma\left(2n,\kappa\right)$ \citep{1991A&A...249...99C}. The special case of $n = 1$ corresponds to the exponential profile often used to describe galactic disks \citep{1970ApJ...160..811F}, and $n = 4$ is the classic canonical profile of an elliptical galaxy \citep{1948AnAp...11..247D}. Initial guesses of the magnitudes come from the Kron aperture photometry, and for the other parameters (source location, $R_e$, position angle $\Theta$, and axial ratio $q$) they are available from the source detection procedure.

As in \citet{2013MNRAS.430..330H}, $\Theta$ and $q$ are held constant with wavelength, while the magnitude of the galaxy is free to vary; this is realized in {\tt GALFITM} by setting the Chebyshev polynomial to a maximum order of 6. The global profile and size of both early-type and late-type galaxies change smoothly and systematically with wavelength, with $n$ increasing and $R_e$ decreasing systematically from the UV to the near-infrared, a consequence of gradients in dust attenuation, stellar population, and metallicity \citep{2012MNRAS.421.1007K}. The wavelength dependence can be described largely by a linear function, although in some cases a second-order function is needed to account for a mild curvature \citep{2013MNRAS.430..330H}. In view of the broad spectral coverage of NIRCam, we allow $n$ and $R_e$ to vary quadratically with wavelength. This should suffice. \citet{2023ApJ...942L..28T} show that the morphology of Lyman-break galaxies at $z > 7$ does not change significantly from the rest-frame UV to the optical.

\begin{figure}[t]
\center{\includegraphics[width=1\linewidth]{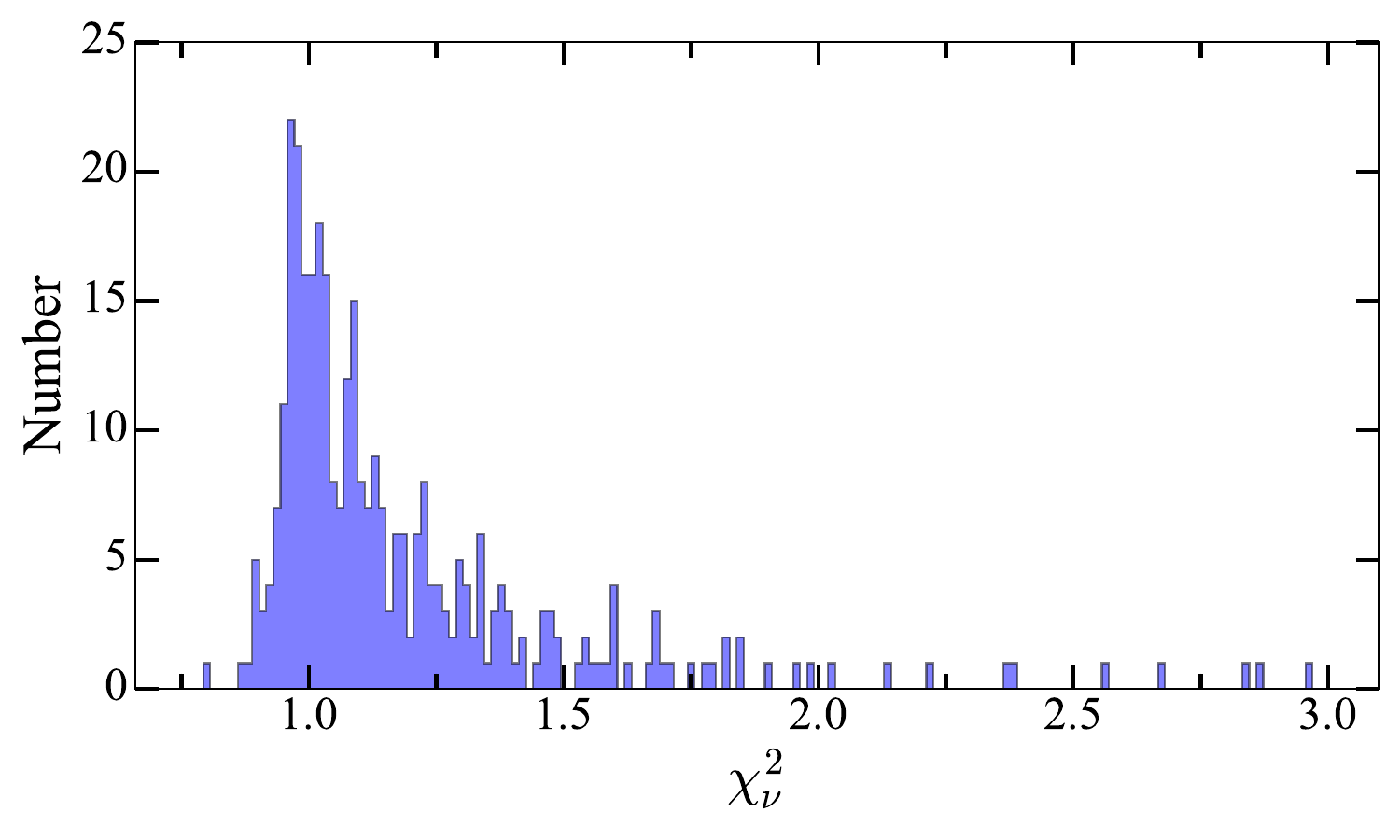}}
\caption{Distribution of $\chi^2_{\nu}$ of the {\tt GALFITM} fits. There are 11 sources with $\chi^2_{\nu}>3$ not shown. \label{fig:chi2_distri}}
\end{figure}

\begin{figure*}[ht]
\begin{center}
\includegraphics[width=1.0\textwidth]{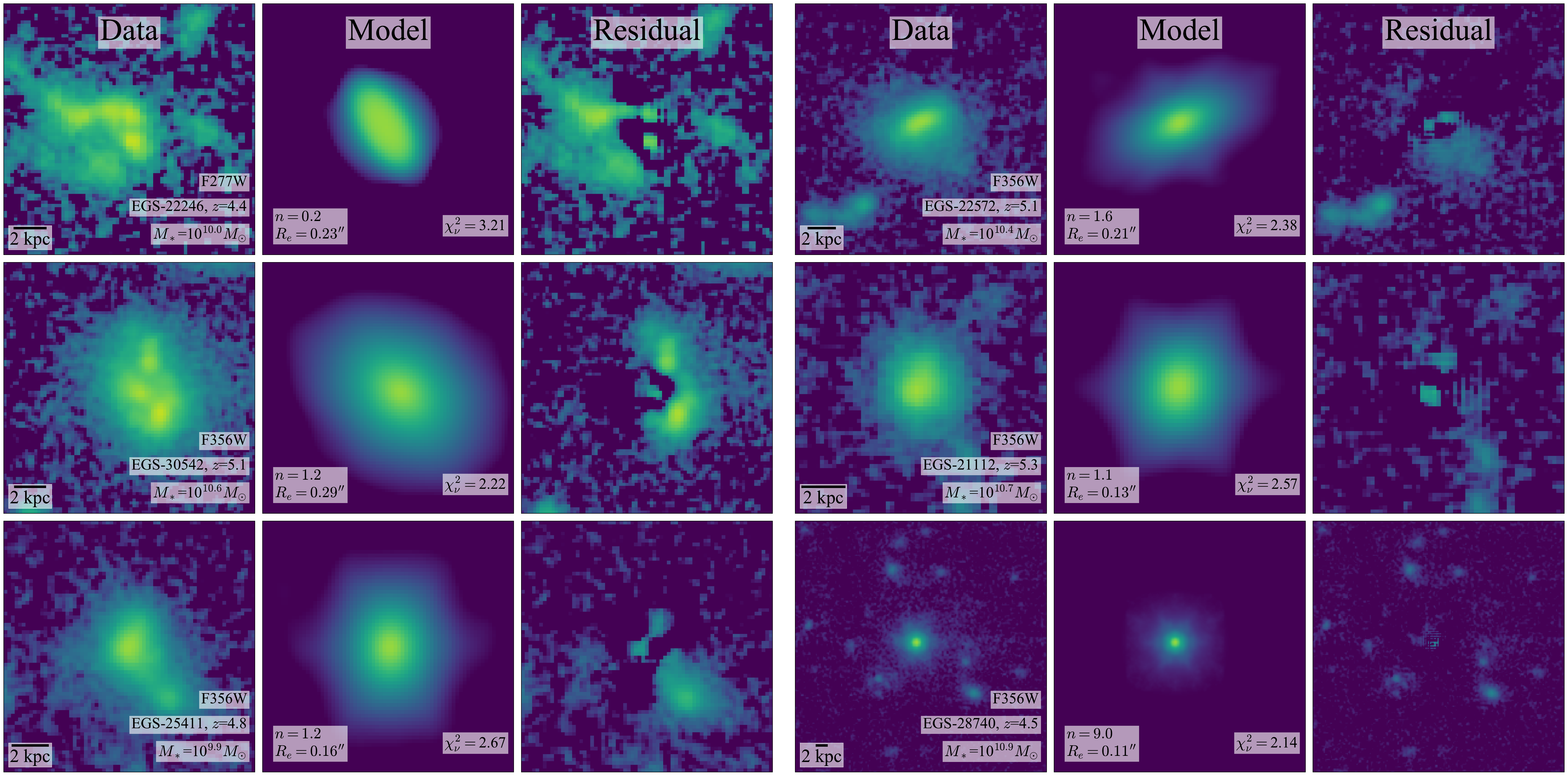}
\end{center}
\caption{As with Figure~\ref{fig:multi_ellip}, but for galaxies whose fits have large residuals. \label{fig:badfits}}
\end{figure*}

\subsection{Results} \label{subsec:results}

Figure~\ref{fig:31079_ellip} shows the simultaneous, multiband fitting results of a sample galaxy. The smooth residuals and $\chi^2_{\nu} \approx 1$ indicate a successful fit. Figure~\ref{fig:multi_ellip} gives additional examples of model fits for the band closest to the rest-frame optical for representative galaxies with disky morphologies ($n = 0.8-1.4$), while Figure~\ref{fig:multi_ellip_1} highlights what might be deemed more spheroid-dominated systems ($n = 2.3-7.9$). Again, the fits are largely satisfactory, as evidenced by the distribution of $\chi^2_{\nu}$ for the entire sample, which is strongly clustered near 1 (Figure~\ref{fig:chi2_distri}). Twenty-three sources exhibit notably higher levels of substructure in their residual maps. These sources, which typically have $\chi^2_{\nu}> 1.9$, are flagged as ``quality 2'' in Table~\ref{tab:all_par}. The more pronounced residuals in these outliers arise from various causes, most commonly because of the presence of complex, multiple components apparently associated with mergers, tidal features, or internal clumpy substructure (Figure~\ref{fig:badfits}). A few have an unusually large S\'ersic $n$ but small $R_e$, suggestive of the presence of an exceptionally compact central component, possibly associated with an active nucleus (C.~H. Chen et al., in preparation). The following discussion (Section~\ref{sec:results}) uses all the galaxies with a quality flag = 1 ($\chi^2_{\nu}\leq 1.9$), representing 93\% of the total sample. The summary of the fitting results in Table~\ref{tab:all_par} pertains to measurements made in the filter closest to the rest-frame optical ($\sim 3800-7800$~\AA) for each galaxy.

\begin{figure}[b]
\begin{center}
\includegraphics[width=0.47\textwidth]{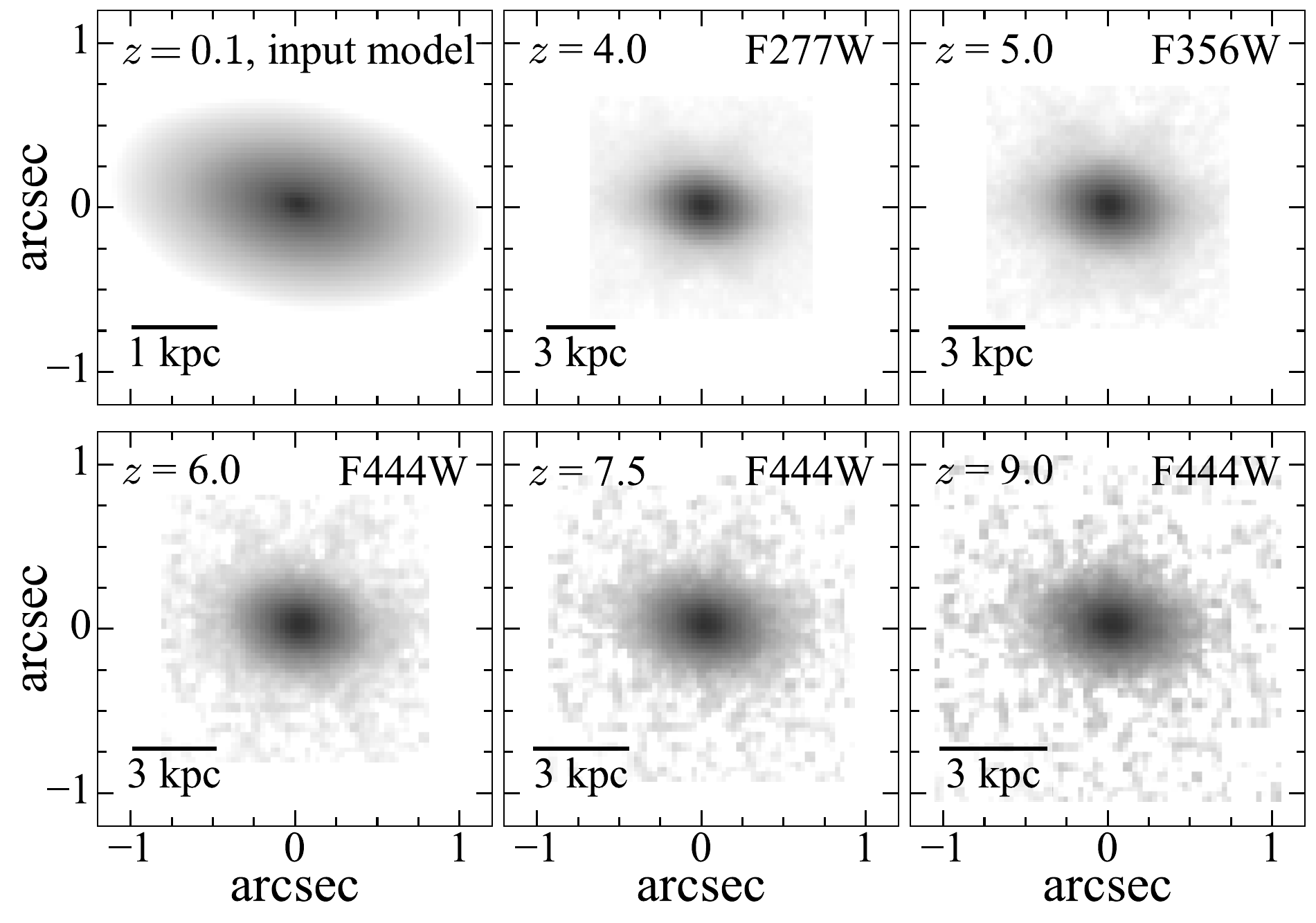}
\end{center} 
\caption{Illustration of a galaxy simulated at different redshifts from $z = 4$ to 9, as observed with JWST/NIRCam in filters that trace the rest-frame optical band, as given in the upper-right corner of each subpanel. \label{fig:mockz_exp}}
\end{figure}

\subsection{Effects of Redshift on the Determination of Galaxy Structure} \label{subsec:z}

The appearance of a galaxy depends on the observing conditions of the telescope and redshift owing to cosmological dimming, resolution effects, and sensitivity \citep[e.g.,][]{1996AJ....112..369G, 1997AJ....114.1741H, 2003ApJS..147....1C, 2008ApJS..175..105B, 2013MNRAS.435..623V, 2016ApJ...824..112D}. We investigate how these observational effects influence our galaxy structure measurements by creating mock images that simulate the same galaxy located at different redshifts. By analyzing the simulated images in exactly the same manner as the real images, we can quantify the differences between the measured and input values of the structural parameters, thereby obtaining a quantitative estimate of the systematic bias and uncertainty contributing to the final error budget of our measurements (Appendix \ref{appen:zeffect}).

We begin by using {\tt GALFIT} to generate S{\'e}rsic model images with known input parameters covering those measured for our real galaxy sample. To convert the model image of a galaxy of a given physical size from low to high redshift, we modify its angular size with redshift as

\begin{equation}
a\propto d_L^{-1}(1+z)^2, \label{eq:2}
\end{equation}

\noindent
where $d_L$ is the luminosity distance. For a given absolute magnitude, the observed surface brightness $I$ changes with redshift as

\begin{equation}
I\propto (1+z)^{-4}. \label{eq:3}
\end{equation}

\noindent
We calculate the image rebinning factor and flux scaling factor following Section~3.1 of \citet{2008ApJS..175..105B}. To convert model images of galaxies at lower redshift to higher redshifts to mimic the CEERS observations, we create mock images with an output pixel scale of $0\farcs030$, convolve them with the PSF for the corresponding filter (Section~\ref{subsec:psf}), introduce realistic noise levels, and add background. As in \citet{2008ApJS..175..105B}, we compute a convolution kernel to produce the NIRCam PSF from the input PSF used for generating the model image. Because the FWHM and geometric shrinking of the input PSF vary according to both the input and target redshift, the convolution kernel is calculated for each input galaxy and output redshift. We convolve the image from the last step with the PSF-matching kernel so that the PSF of the final image matches that of the desired NIRCam band. Noise in the output image mainly comes from (1) random Poisson noise associated with the galaxy flux, and (2) correlated noise and intrinsic variations of the background, which we simulate using blank regions from actual NIRCam images.

Starting at $z = 0.1$, we generate a series of mock galaxies at increasing redshifts, focusing on the interval $4 < z < 9$ for which we divide into 11 discrete redshift bins separated by $\Delta z = 0.5$. The increase of angular size with redshift (Equation~\ref{eq:2}) and the concomitant dimming of surface brightness (Equation~\ref{eq:3}), coupled with the observational characteristics of NIRCam, underscore the main trends visible in the simulated images (Figure~\ref{fig:mockz_exp}). Our immediate interest primarily concerns the rest-frame optical band appropriate to each of the sources of our sample (see Appendix \ref{appen:zeffect} for details).

\begin{figure*}[t]
\gridline{\fig{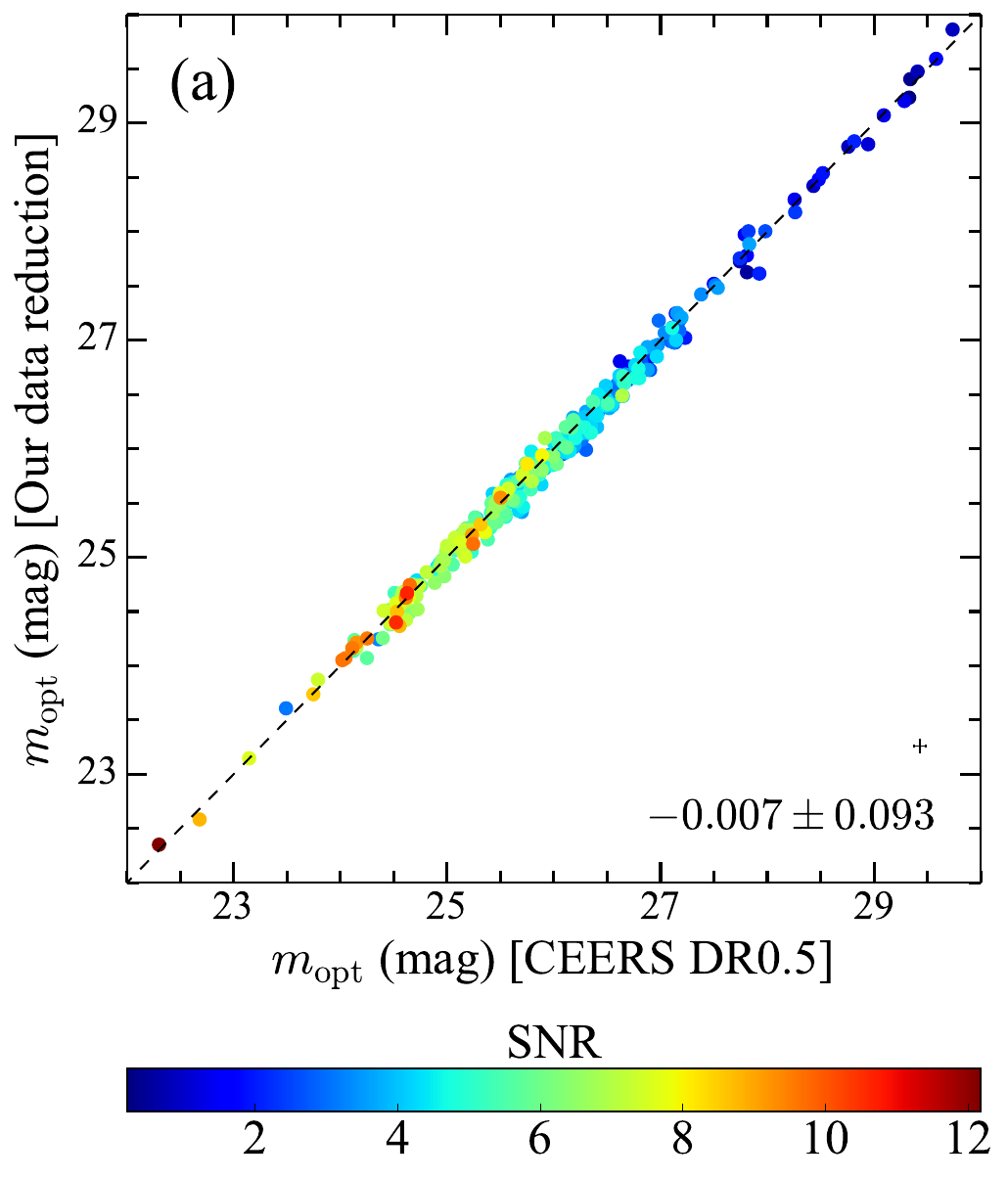}{0.3\textwidth}{}
          \fig{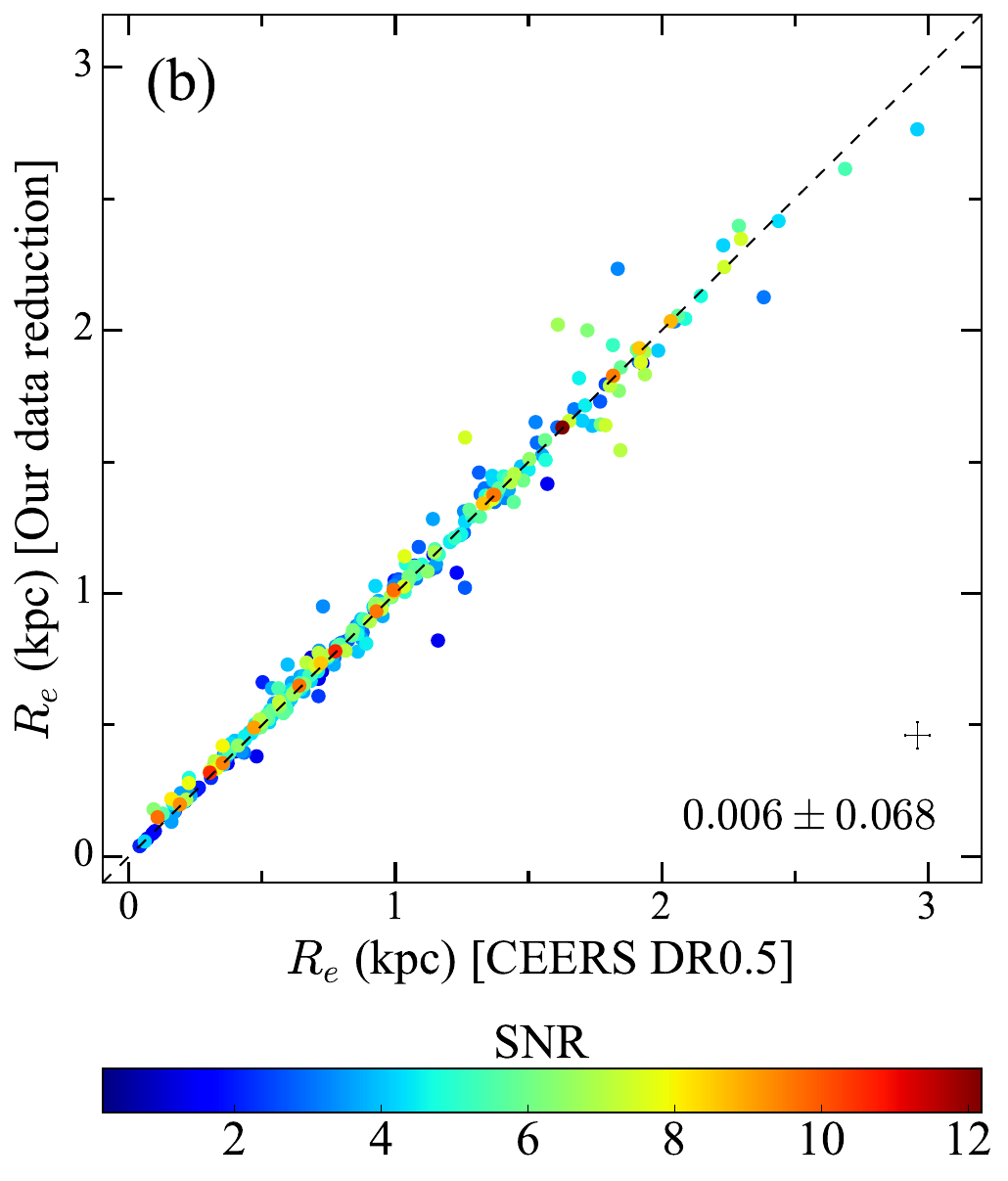}{0.3\textwidth}{}
          \fig{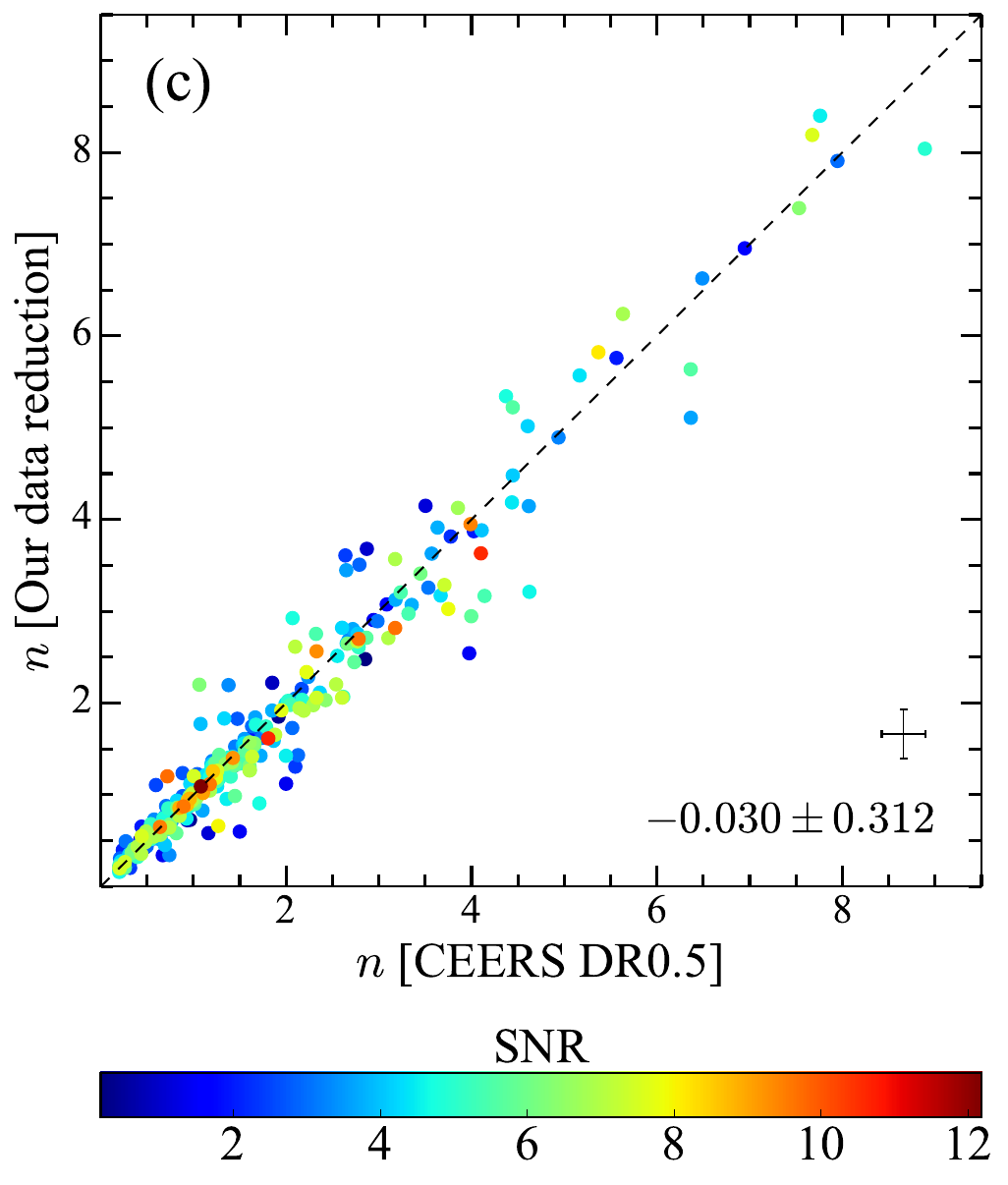}{0.3\textwidth}{}
          }
\caption{Comparison between the results derived using images from the first public release (Data Release~0.5) of CEERS and from those reduced by us, which have a finer pixel scale after drizzling ($0\farcs 015\,{\rm pixel}^{-1}$ instead of $0\farcs 030\,{\rm pixel}^{-1}$), for the (a) magnitude, (b) effective radius, and (c) S\'ersic index derived from the filter that most closely approximates the rest-frame optical band. Median differences (y-axis $-$ x-axis) and standard deviations are given in the lower-right corner of each panel, which also shows the typical uncertainties. The dashed line indicates the 1:1 relation. Points are color-coded by the signal-to-noise ratio in the filter that most closely approximates the rest-frame optical band. \label{fig:pc}}
\end{figure*}

\subsection{Influence of the Pixel Scale} \label{appen:pixelscale}

To investigate the potential impact of the output pixel scale after drizzling on our measured results, we repeat the model fits with the images reduced by ourselves using a finer pixel scale. As {\tt GALFITM} requires a common pixel scale across all filters, we reproject the mosaics of the three SW filters to the pixel scale of the LW filters ($0\farcs 030$) prior to fitting. Judging by the median difference and standard deviation of the two sets of measurements (Figure~\ref{fig:pc}), we conclude that the effect is minimal on $m_{\mathrm{opt}}$ ($-0.007\pm 0.093$ mag) and $R_e$ ($0.006\pm 0.068$~kpc), but the scatter for $n$ is nonnegligible ($-0.030\pm 0.312$). The galaxies that exhibit the worst scatter in $n$ tend to have lower signal-to-noise ratios. The final error budget of each measured parameter (Table~\ref{tab:all_par}) incorporates the uncertainty introduced by the choice of pixel scale used in the drizzling process.

\subsection{Error Budget} \label{subsec:error}

The rich structural complexities of galaxies are difficult to capture with simple parametric models. Parameters returned by codes such as {\tt GALFIT} or {\tt GALFITM} may suffer from various degrees of systematic bias and degeneracy that can underestimate the real uncertainties, even if the formal statistical errors are small and the fitting residuals look acceptable \citep[see, e.g.,][]{2007ApJS..172..615H, 2008ApJS..179..283K, 2013MNRAS.435..623V, 2017ApJ...845..114G, 2021ApJ...911...94Z, 2022ApJ...934..130Z}. We design realistic input-output experiments to determine the real parameter uncertainties using a set of mock galaxies generated from the best-fit parameters of each galaxy in our sample. Following \citet{2022ApJ...934..130Z}, we use the best-fit parameters derived for each galaxy to construct mock images that exactly mimic the object-specific parameters of the actual observation. For each galaxy in each band, we generate 100 realizations of mock observations that account for the Poisson noise associated with the source, background Gaussian noise, and the properties of the specific image (gain, exposure time, and background variation). Then, we use {\tt GALFITM} to repeat the model fitting and adopt the median value and standard deviation of the 100 results as the measurement and its error. 

As discussed in Section~\ref{subsec:DataReduction} and documented in Section~\ref{appen:pixelscale}, our final choice of drizzling the images to a common scale of $\sim 0\farcs 030\,{\rm pixel}^{-1}$ has little impact on $m$ and $R_e$, but the effect is not negligible for $n$. We take the difference between the measurements based on images reduced with two different pixel scales as another source of uncertainty. This, along with the contribution from redshift effects and the standard statistical uncertainty returned by {\tt GALFITM}, constitute the quadrature contributions to the final error budget of each parameter.

\begin{figure}[t]
\center{\includegraphics[width=1.0\linewidth]{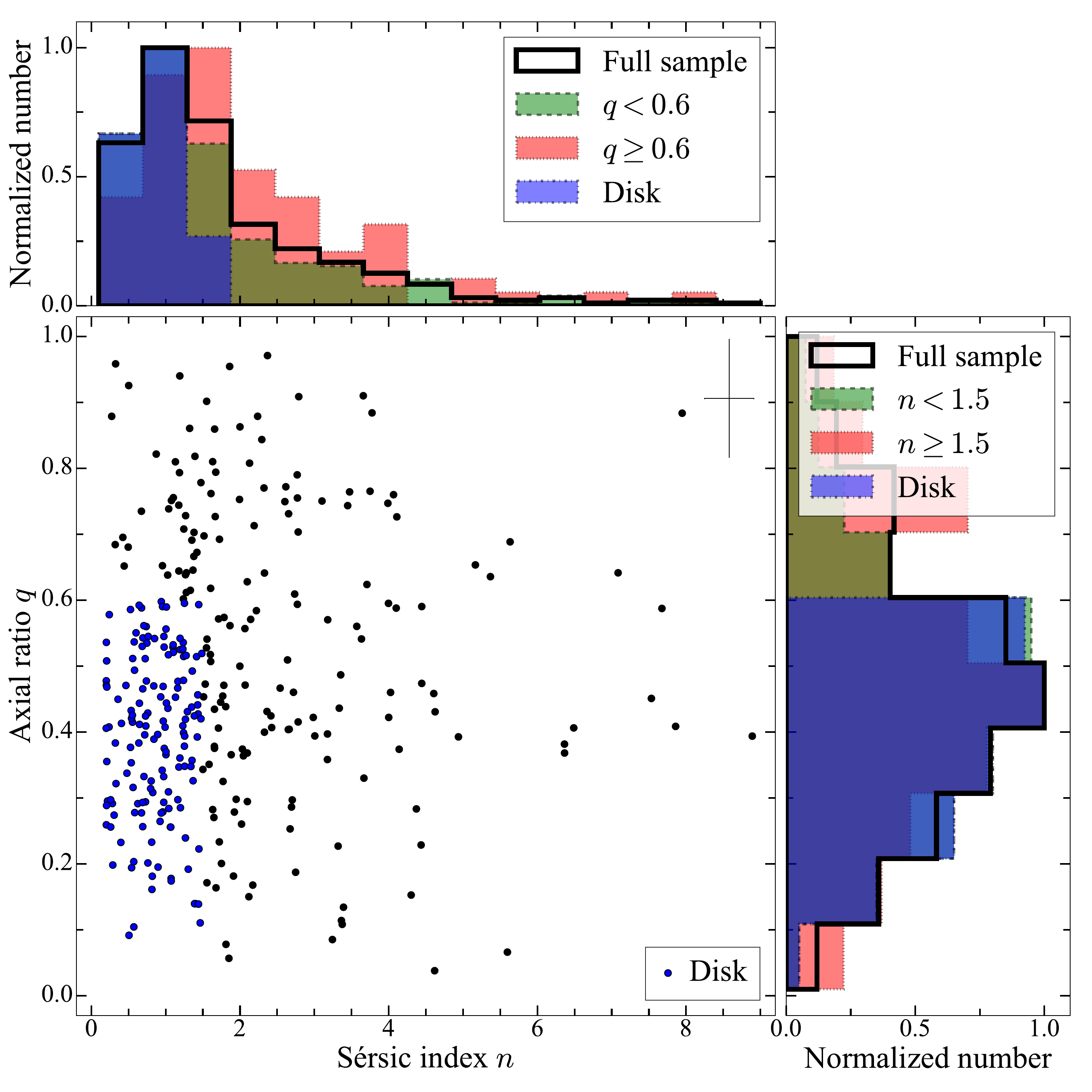}}
\caption{The normalized distribution of axial ratio $q$ and S\'ersic index $n$ measured in the rest-frame optical. Typical uncertainties are shown in the upper-right corner of the main panel. The histograms highlight in green objects that have either $q < 0.6$ or $n < 1.5$, and in red those that have either $q \geq 0.6$ or $n \geq 1.5$. Sources characterized by both $q < 0.6$ and $n < 1.5$ constitute the most conservative candidates for disk-dominated galaxies.
\label{fig:n_distri}}
\end{figure}

\begin{figure*}[ht!]
\centering
\includegraphics[width=0.48\textwidth]{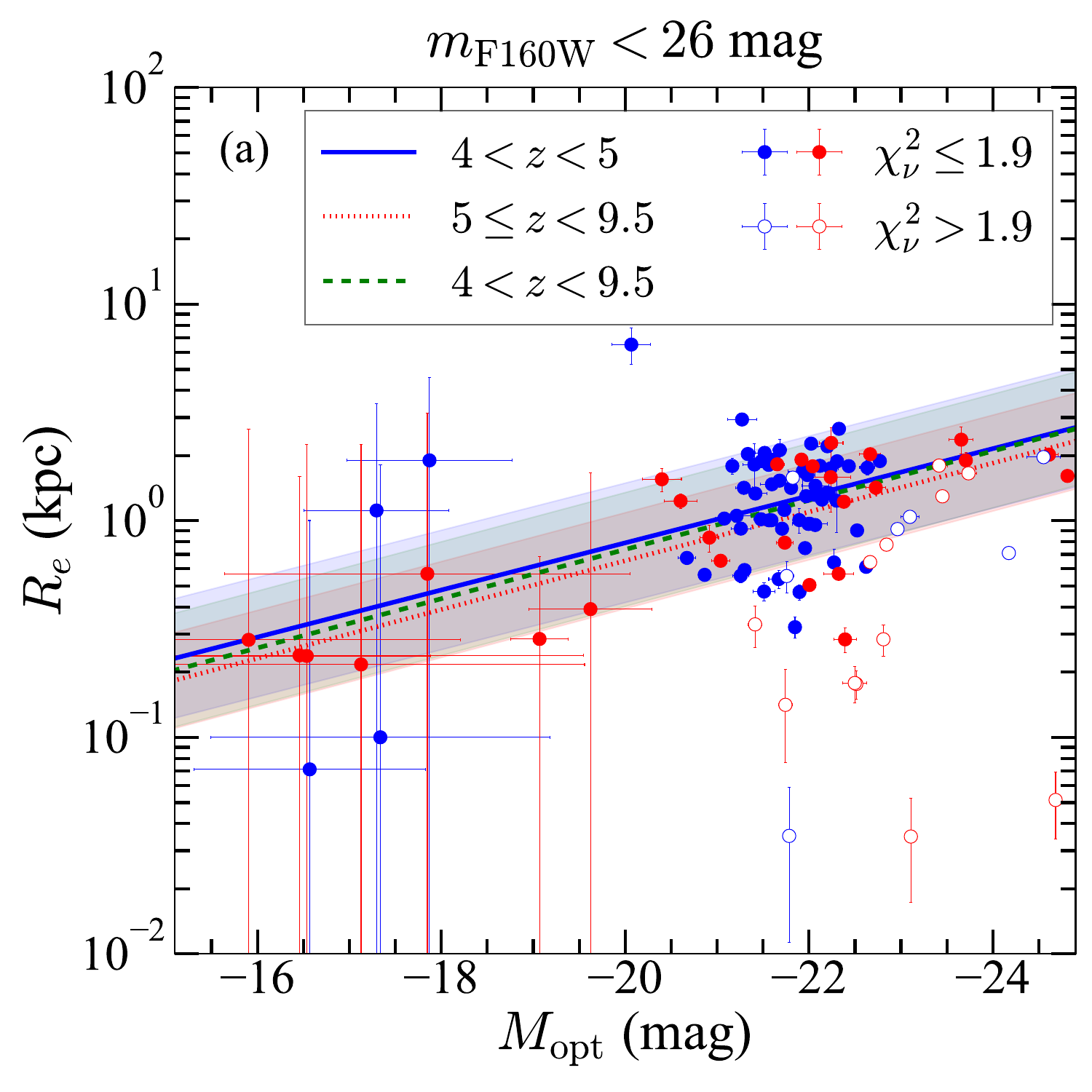}
\includegraphics[width=0.48\textwidth]{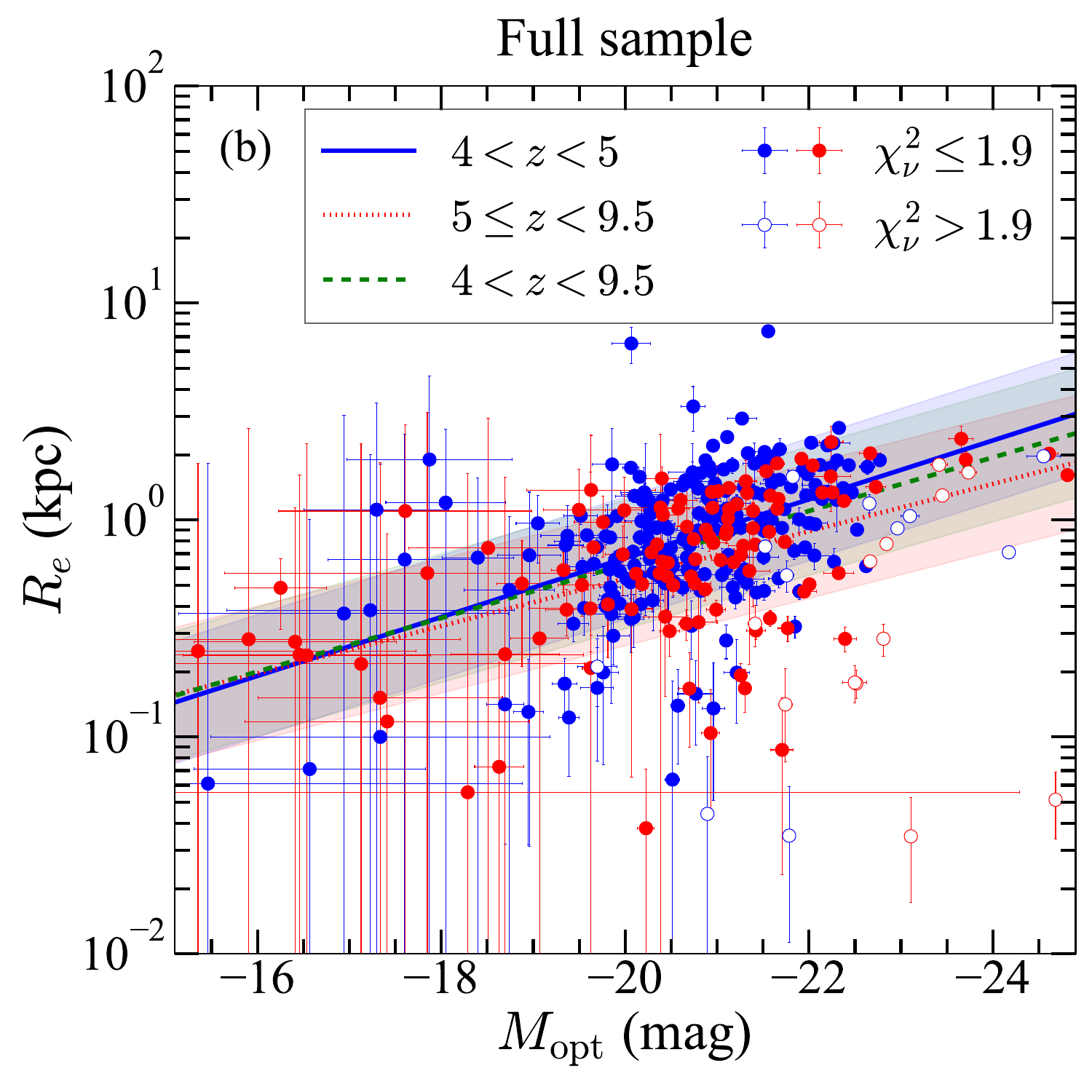}
\caption{Luminosity-size relation for galaxies at $4 < z < 5$ (blue points, blue solid line), $5 \leq z < 9.5$ (red points, red dotted line), and $4 < z < 9.5$ (green dashed line), for (a) the subsample of 84 galaxies with $m_{\mathrm{F160W}} < 26$~mag and (b) the full sample of 347 galaxies; the shaded region corresponds to the log-normal dispersion of $\log R_e$ (Table~\ref{tab:m_re}). Filled and open points correspond to fits with $\chi^2_{\nu}\leq 1.9$ and $\chi^2_{\nu} > 1.9$, respectively. The absolute magnitudes and effective radii pertain to the filter that most closely approximates the rest-frame optical band.
\label{fig:mass_re}}
\end{figure*}

\section{Discussion} \label{sec:results}

\subsection{Incidence of Galactic Disks}

The distribution of S{\'e}rsic indices for the full sample at $4 < z < 9.5$ strongly peaks at $n \approx 1$, with a long tail that extends to values as large as $n \approx 8$ (Figure~\ref{fig:n_distri}). In total, $\sim 55\%$ of the sample are well fit with $n < 1.5$, an often-invoked criterion to designate disk-dominated systems. The residuals of the fits are relatively clean (Figure~\ref{fig:multi_ellip}), indicating that a simple, single-component S\'ersic model suffices to describe the overall light distribution. The choice of pixel scale for drizzling (Section~\ref{appen:pixelscale}) and redshift effects (Appendix~\ref{appen:zeffect}) can increase the uncertainty of individual measurements of $n$, but these factors do not bias the overall distribution. The axial ratio is more challenging to interpret for a randomly oriented mixture of galaxies of different morphological types. Disk-dominated systems exhibit axial ratios skewed toward lower values compared to ellipticals, which tend to be rounder with axial ratios closer to 1 \citep[e.g.,][]{Padilla+2008}. If we designate galaxies with $q < 0.6$ as having a substantial disk component, the disk fraction is $\sim 75\%$. If, on the other hand, we conservatively require that the galaxy has both $n < 1.5$ and $q < 0.6$, then we can place a firm lower limit of $\sim 45\%$ on the incidence of galactic disks in our sample. In agreement with other recent analyses of high-redshift galaxies based on CEERS \citep{2022ApJ...938L...2F, 2022arXiv221001110F, 2022arXiv220806516J, 2022arXiv220801630N, 2023ApJ...946L..15K, 2023ApJ...942L..42R}, the incidence of disky galaxies at this early cosmic epoch is much higher than expected based on previous HST studies \citep[e.g.,][]{2015ApJS..221...11K}. Differences in the depth and wavelength coverage between JWST and HST may contribute to this apparent discrepancy in disk fraction detected from the two missions. The deeper images of JWST can detect more easily the faint disk component previously missed by HST. Conversely, a galaxy may be mistaken as a pure spheroid if HST is only sensitive to its compact central component. An intrinsically disky galaxy with UV-bright star-forming clumps may also be misclassified as irregular if HST misses the underlying disk component.

We caution that without kinematical information, image-based analyses, including ours, can overestimate the fraction of disk-dominated systems. \citet{Vega-Ferrero+2023} apply self-supervised machine learning to explore the morphological diversity of galaxies at $z\geq 3$. By comparing traditional morphological measures with the physical parameters obtained from cosmological simulations, they find that approximately 50\% of the galaxies visually classified as disks based on the imaging data are intrinsically prolate or spheroidal objects.

\subsection{Luminosity-size Relation}

The correlation between the luminosity of a galaxy and its spatial distribution within the galaxy, described by the luminosity-size relation, offers important insights into the evolution and assembly history of the galaxy. Figure~\ref{fig:mass_re} shows the distribution of rest-frame optical luminosity versus effective radius, for the subset of 84 galaxies with $m_{\mathrm{F160W}} < 26$~mag, which is the 90\% detection completeness limit for extended sources in the catalog of \citet[][see their Section~5.1]{2017ApJS..229...32S}, as well as for the entire sample of 347 galaxies. We further divide each sample into two redshift bins to see whether there is any evolution between $4 < z < 5$ and $5 \leq z < 9.5$. Correlations are evident for all the samples considered.

To quantify the observed empirical relations, we assume that the size distribution at fixed $L$ follows a log-normal function \citep{2003MNRAS.343..978S}, a prescription motivated by the disk formation theory of \citet{1980MNRAS.193..189F}. Describing the luminosity-size relation by a power law \citep[e.g.,][]{2013ApJ...765...68H}, the probability density function of the $L-R_e$ pair can be expressed as

\begin{equation}
\begin{aligned}
P\left(R_e, L; \beta, R_0, \sigma\right)=&\frac{1}{R_e\sigma\sqrt{2\pi}}\times \\
&\exp{\left(-\frac{\left(\ln R_e-\ln \overline{R_e}\right)^2}{2\sigma^2}\right)},
\end{aligned}
\end{equation}

\noindent
with

\begin{equation}
\overline{R_e}=R_0\left(\frac{L}{L_0}\right)^{\beta},
\end{equation}

\noindent
where $\overline{R_e}$ is the peak of the size distribution, $\sigma$ is the log-normal dispersion of $\log R_e$, $L_0$ is the characteristic luminosity corresponding to an absolute magnitude $M = -21.0$, $R_0 = \overline{R_e}$ at $L_0$, and $\beta$ is the slope of the relation. Table~\ref{tab:m_re} lists the best-fit parameters. 

Within the considerable scatter of $\sim 0.5-0.6$~dex, the fits of both redshift ranges are statistically consistent with each other, and they are not strongly dependent on whether we apply the completeness cut of \citet{2017ApJS..229...32S}. Our results for the redshift bins $4<z<5$ and $5\leq z<9.5$ agree with those of studies at similar redshifts based on the rest-frame optical/UV \citep[e.g.,][]{2015ApJS..219...15S, 2022ApJ...938L..17Y}, indicating little evidence for wavelength dependence of the luminosity-size relation at $4<z<9.5$. The higher redshift bin may have a marginally lower zero point: for the $m_{\mathrm{F160W}} < 26$~mag sample, $R_0 = 0.85 \pm 0.08$~kpc at $5 \leq z<9.5$, to be compared with $R_0 = 1.02 \pm 0.09$~kpc for $4 < z < 5$; the corresponding values for the full sample are $R_0 = 0.69 \pm 0.05$~kpc and $R_0 = 0.91 \pm 0.04$~kpc, respectively. These results are consistent trends reported by \cite{2015ApJS..219...15S} for the rest-frame UV band at $0< z< 8$.

\begin{deluxetable}{ccccc}
\tabletypesize{\scriptsize}
\tablewidth{0pt} 
\tablecaption{Best-fit Parameters of the Luminosity-size Relation\label{tab:m_re}}
\tablehead{
\colhead{Sample} & 
\colhead{$z$} & 
\colhead{$\beta$} & 
\colhead{$R_0$} & 
\colhead{$\sigma$}\\
\colhead{} & \colhead{} & \colhead{}& \colhead{(kpc)}& \colhead{(kpc)}}
\colnumbers
\startdata
{}&$4< z<5$       & $0.27 \pm 0.07$ & $1.02\pm 0.09$ & $0.63\pm 0.06$ \\
(a)&$5\leq z< 9.5$ & $0.28 \pm 0.04$ & $0.85\pm 0.08$ & $0.51\pm 0.07$ \\
{}&$4< z< 9.5$ & $0.28 \pm 0.04$ & $0.96\pm 0.06$ & $0.61\pm 0.05$ \\
\hline
{}&$4< z<5$       & $0.34 \pm 0.04$ & $0.91\pm 0.04$ & $0.65\pm 0.03$ \\
(b)&$5\leq z< 9.5$ & $0.27 \pm 0.04$ & $0.69\pm 0.05$ & $0.71\pm 0.05$ \\
{}&$4< z< 9.5$ & $0.31 \pm 0.03$ & $0.83\pm 0.03$ & $0.68\pm 0.03$ \\
\enddata
\tablecomments{Col. (1): Sample, where (a) corresponds to the subsample of 84 galaxies with $m_{\mathrm{F160W}} < 26$~mag and (b) denotes the full sample of 347 galaxies.  Col. (2): Redshift range. Col. (3): Slope. Col. (4): Intercept. Col. (5): Log-normal dispersion of $\log R_e$.}
\end{deluxetable}

\begin{figure*}[ht!]
\centering
\includegraphics[width=0.48\textwidth]{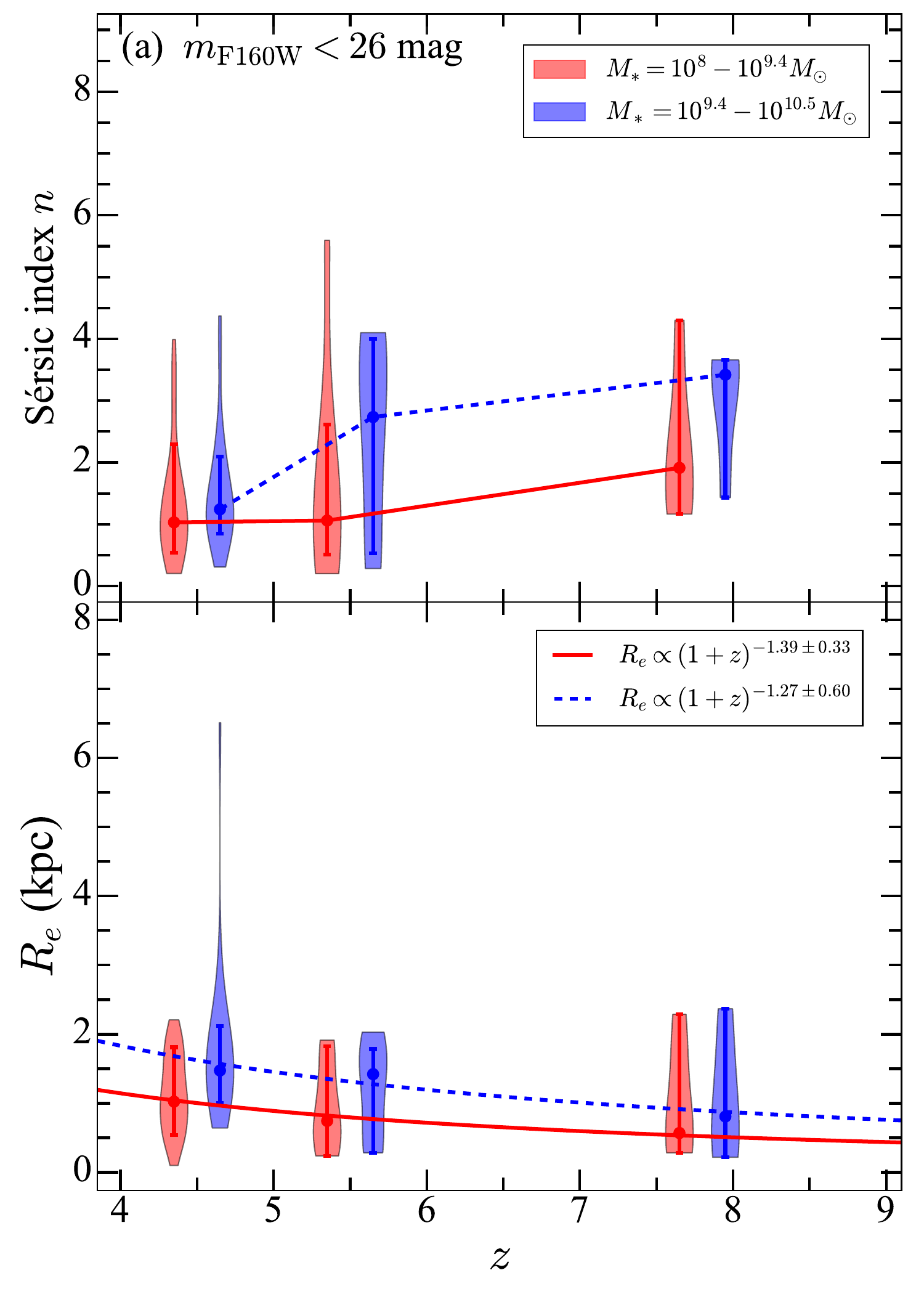}
\includegraphics[width=0.48\textwidth]{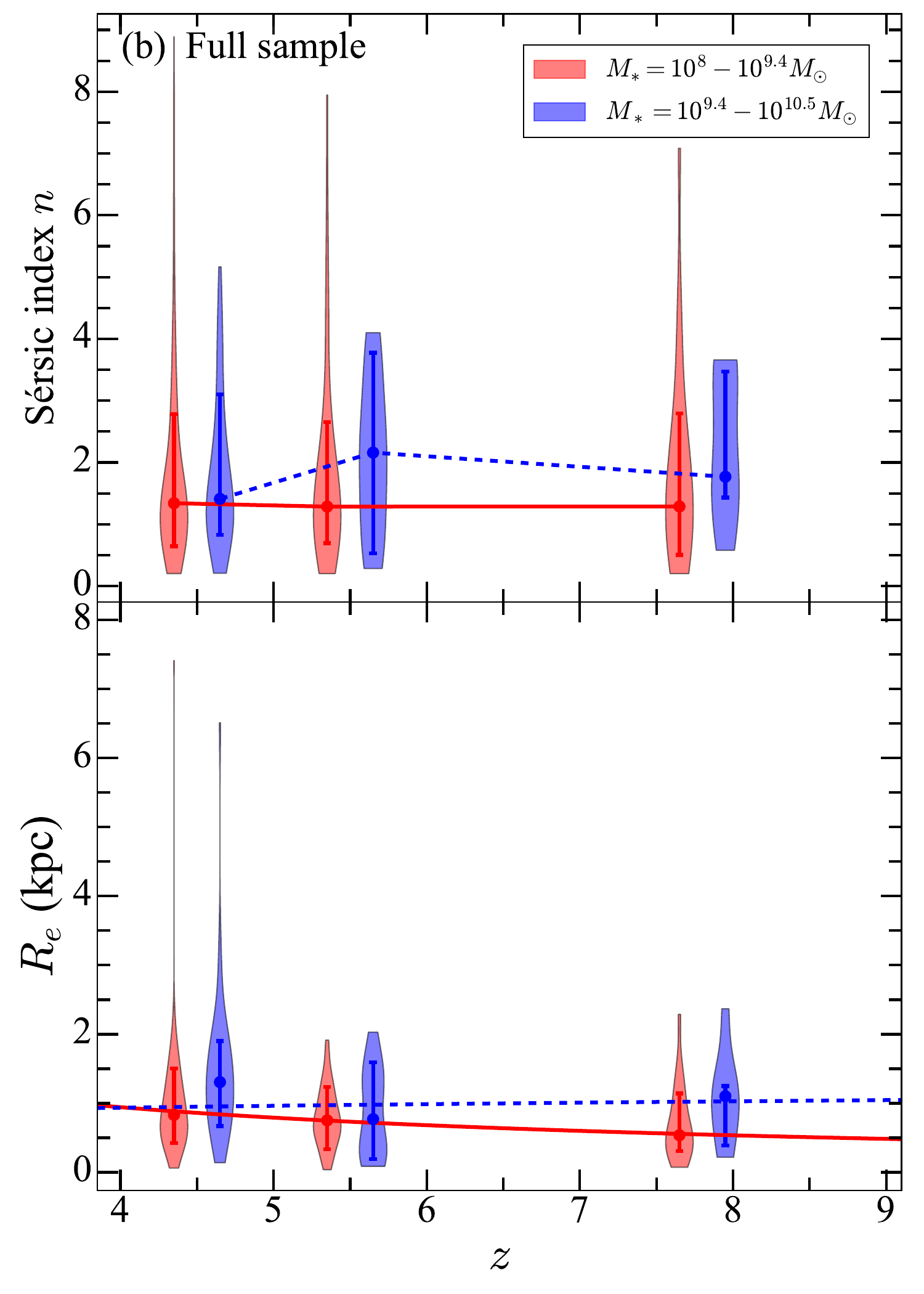}
\caption{Evolution with the redshift of S{\'e}rsic index $n$ (top) and effective radius $R_e$ (bottom) for two stellar mass bins (red: $M_* = 10^{8}-10^{9.4}\, M_{\odot}$; blue: $M_* = 10^{9.4}-10^{10.5}\, M_{\odot}$), shown separately for (a) the subsample of 84 galaxies with $m_{\rm F160W} < 26$~mag and (b) the full sample. The error bars define the 15\% and 85\% percentile of the distribution for each redshift bin, whose violin-style shaded region displays the probability density distribution smoothed by a Gaussian kernel and normalized by the number of galaxies in each bin; the relative width of the violin shape in each bin corresponds to the fraction of galaxies with that value of $n$ or $R_e$. \label{fig:n_re_z}}
\end{figure*}

\subsection{Evolution of S{\'e}rsic Index and Effective Radius}

The evolutionary pathway of a galaxy, and hence its structural parameters, may depend on mass. To discern the possible evolution of galaxy structure with redshift, we focus on two stellar mass bins using sources that overlap in the redshift range $4 < z < 9.5$: $M_* = 10^{8}-10^{9.4}\, M_{\odot}$ (233 galaxies) and $M_* = 10^{9.4}-10^{10.5}\, M_{\odot}$ (73 galaxies). To warrant against sample incompleteness, we further isolate the subset of galaxies with $m_{\mathrm{F160W}} < 26$~mag that meet the 90\% detection completeness limit of \citet{2017ApJS..229...32S}, at the expense of drastically reducing the sample to merely 46 galaxies for the low-mass bin and 38 galaxies for the high-mass bin.

Focusing first on the subset of galaxies with $m_{\mathrm{F160W}} < 26$~mag, Figure~\ref{fig:n_re_z}a (top panel) suggests that $n$ drops systematically from $z\gtrsim 6$ to $z\approx 4$. If we adopt a fiducial criterion of $n<1.5$ to designate a disky morphology, the fraction of disk-dominated galaxies, for both mass bins combined, roughly doubles from $\sim 30\%$ at $z = 6-9.5$ to $\sim 60\%$ at $z = 4-6$. Our results qualitatively agree with those of \cite{2022arXiv221001110F} and \cite{2023ApJ...946L..15K} based on visual classifications. However, for the low-mass and high-mass bin, the difference between the median values of $n$ at $z\gtrsim 6$ and $z = 4-6$ is only 0.55$\sigma$ and 1.45$\sigma$, respectively (the $x\sigma$ difference is calculated as $x=\left|m_1-m_2\right|/\sqrt{\sigma^2_1+\sigma^2_2}$, where $m_1$ and $m_2$ denote the median value of the distribution, and $\sigma_1$ and $\sigma_2$ give the standard deviation of the distribution). According to the two-sample Kolmogorov-Smirnov test, we cannot reject the null hypothesis that the distributions of $n$ at $z\gtrsim 6$ and $z = 4-6$ are similar, with $p-$value of 0.1 for both the low-mass and high-mass bins, suggesting that apparent redshift evolution of $n$ is insignificant.

Galaxy effective radius also seems to exhibit a mild but systematic increase toward lower redshift (Figure~\ref{fig:n_re_z}a, bottom panel), from a median $R_e \approx 0.57$~kpc and $0.81$~kpc for the low-mass and high-mass bin at $z = 6-9.5$, to corresponding values of $R_e \approx 1.02$~kpc and $1.46$~kpc at $z = 4-6$. However, for the low-mass and high-mass bin, the difference between the median values of $R_e$ at $z\gtrsim 6$ and $z = 4-6$ is only 0.47$\sigma$ and 0.49$\sigma$, respectively. The two-sample Kolmogorov-Smirnov test cannot reject the null hypothesis that the distributions of $R_e$ of the two redshift intervals are similar, with $p-$value of 0.52 and 0.34 for the low-mass and high-mass bin, respectively. As mentioned in Section~\ref{sec:intro}, the expected size scaling from semi-analytical models is $R_e \propto H(z)^{-1}$ at a fixed halo circular velocity or $R_e \propto H(z)^{-\frac{2}{3}}$ at a fixed halo mass \citep{1980MNRAS.193..189F, 1998MNRAS.295..319M}, which, for $H(z)\sim(1+z)^{\frac{3}{2}}$ at $z > 2$, translate to $R_e \propto (1+z)^{-\frac{3}{2}}$ and $R_e \propto (1+z)^{-1}$, respectively. As observations track the evolution of $R_e$ at fixed $M_{*}$ (or $L$), we expect to find a trend that falls between these two functional forms \citep{2004ApJ...600L.107F}. Parameterizing the size evolution as $R_e \propto (1+z)^{-\alpha}$, the best-fit relation yields a power-law slope of $\alpha = 1.39\pm0.33$ for the low-mass bin and $\alpha = 1.27\pm0.60$ for the high-mass bin, which are statistically indistinguishable given their large uncertainties but broadly agree with theoretical expectations. These results are consistent with the predictions for the size evolution of galaxies at $3 \leq z \leq 6$ made by \cite{2023ApJ...946...71C} using synthetic images generated from cosmological simulations tailored for JWST observations.

When the entire sample is considered in aggregate (Figure~\ref{fig:n_re_z}b), neither $n$ nor $R_e$ exhibits any significant trend with redshift over the range $z = 4-9.5$. This holds for both the high-mass and low-mass objects. For both mass bins combined, the difference between the median values of $n$ or $R_e$ at $z\gtrsim 6$ and $z = 4-6$ is less than 0.3$\sigma$. These conclusions are further confirmed through the two-sample Kolmogorov-Smirnov test. We are unsure how to interpret these results, except to note that any trends based on small-number statistics at this early stage of the JWST mission should be regarded with extreme caution.

\section{Summary} \label{sec:conclusions}

We study the structural parameters of 347 galaxies at $4<z<9.5$ using $\sim 34.5$ arcmin$^2$ of JWST NIRCam data from the CEERS program covering seven bands from $\sim 1$ to $4.4\,\mu$m. With the aid of {\tt GALFITM}, we perform two-dimensional, simultaneous, multiband model fitting to derive robust seven-band photometry and global structural parameters. After evaluating different methods of PSF construction, we finally derive PSFs from stacking isolated, bright stars. The final error budget of the structural parameters takes into consideration the influence of the pixel scale after drizzling and the effects of cosmological redshift based on realistic mock simulations. 

Our main results are as follows:

\begin{enumerate}

\item We detect a significant population of disk-dominated galaxies. The distribution of global S\'ersic indices in the rest-frame optical band peaks at $n \approx 1$, with $\sim 55\%$ of the sample having $n<1.5$. If, in addition, we require that disky galaxies have an axial ratio $q < 0.6$, we place a conservative lower limit of $\sim 45\%$ on the incidence of galactic disks.

\item Galaxies follow a relation between rest-frame optical luminosity and size, over the entire redshift range of $4<z<9.5$ and separately over the intervals $4 < z < 5$ and $5 \leq z < 9.5$. Galaxies in the higher redshift bin are marginally more compact ($R_e = 0.69 \pm 0.05$~kpc) than those in the lower redshift bin ($R_e = 0.91 \pm 0.04$~kpc). These results are qualitatively consistent with previous findings in the rest-frame UV band.

\item Within the limitations of the current sample size, we find no significant redshift evolution of $n$ or $R_e$ at these early epochs.

\end{enumerate}

Future work can be improved in several directions. Without spectroscopic confirmation, it remains possible that some of the objects in our sample may have inaccurate photometric redshifts. For convenience and consistency with other recent work in the literature, we made use of available photometric redshifts and stellar masses from the catalog of \citet{2017ApJS..229...32S}, which was based on pre-JWST observations. A consistent set of updated photometric redshifts and stellar masses should be derived by incorporating the JWST data. The currently limited sample should be expanded using the remaining CEERS fields, as well as other NIRCam imaging surveys, such as COSMOS-Web \citep{2022arXiv221107865C}, despite having shallower depth and less extensive bandpass coverage. Lastly, it is clear that the NIRCam images are beginning to reveal nascent internal substructures, even for galaxies in the epoch of reionization. More sophisticated analysis is needed to characterize the main structural components in order to elucidate the birth of the Hubble sequence.

\begin{acknowledgments}
We thank the referee for helpful comments and suggestions. This work was supported by the National Key R\&D Program of China (2022YFF0503401), the National Science Foundation of China (11721303, 11991052, 12011540375, 12233001), and the China Manned Space Project (CMS-CSST-2021-A04, CMS-CSST-2021-A06). We are very grateful to Jing Wang, Masafusa Onoue, Yang Li, Dewang Xu, Zhengrong Li, Lulu Zhang, and Fengwu Sun for their assistance.
\end{acknowledgments}

\vspace{5mm}
\facility{JWST (NIRCam)}
\software{{\tt Astropy} \citep{2013A&A...558A..33A,2018AJ....156..123A}, {\tt GALFIT} \citep{2002AJ....124..266P, 2010AJ....139.2097P}, {\tt GALFITM} \citep{2013MNRAS.430..330H, 2013MNRAS.435..623V}, {\tt Matplotlib} \citep{2007CSE.....9...90H}, {\tt NumPy} \citep{2020Natur.585..357H}, {\tt Photutils} \citep{2020zndo...4044744B}, {\tt PSFr} \citep{2022ascl.soft10005B}, {\tt SciPy} \citep{2020NatMe..17..261V}, STScI JWST Calibration Pipeline (\url{jwst-pipeline.readthedocs.io}), {\tt WebbPSF} \citep{2014SPIE.9143E..3XP}}

\newpage
\appendix

\section{Performance of Different PSF Models} \label{appen:psf}

\begin{figure*}[th!]
\figurenum{A1}
\begin{center}
\includegraphics[width=1.0\textwidth]{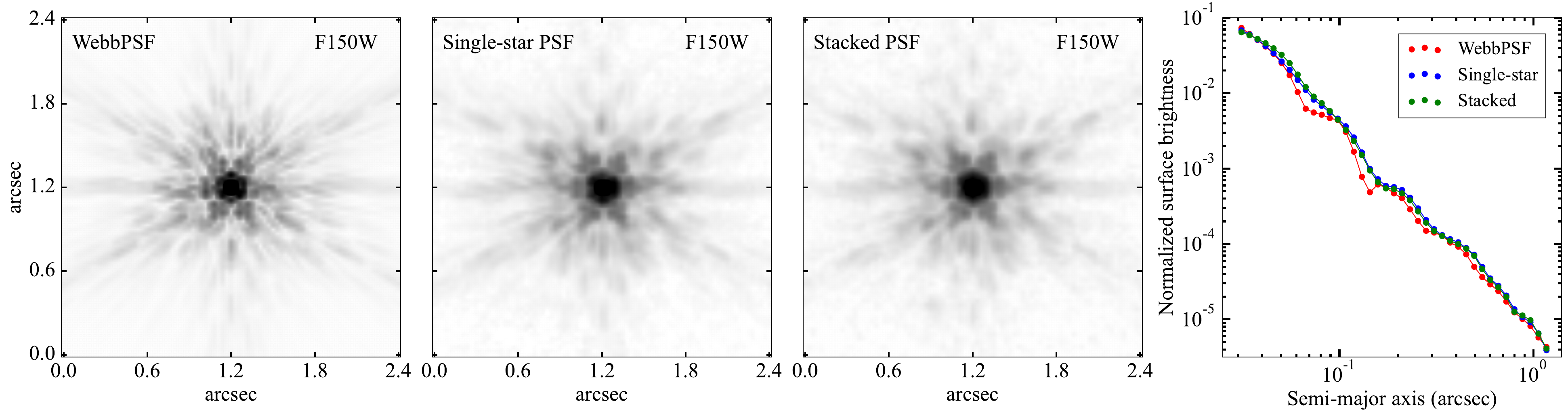}
\end{center}
\caption{Comparison of different PSF models. The images show, from left to right, the PSF in the F150W filter generated from {\tt WebbPSF}, a single star, and stacking multiple stars. The right panel plots the normalized radial surface brightness profile of the three models. \label{fig:psfs}}
\end{figure*}

\begin{figure*}[th!]
\figurenum{A2}
\begin{center}
\includegraphics[width=1.0\textwidth]{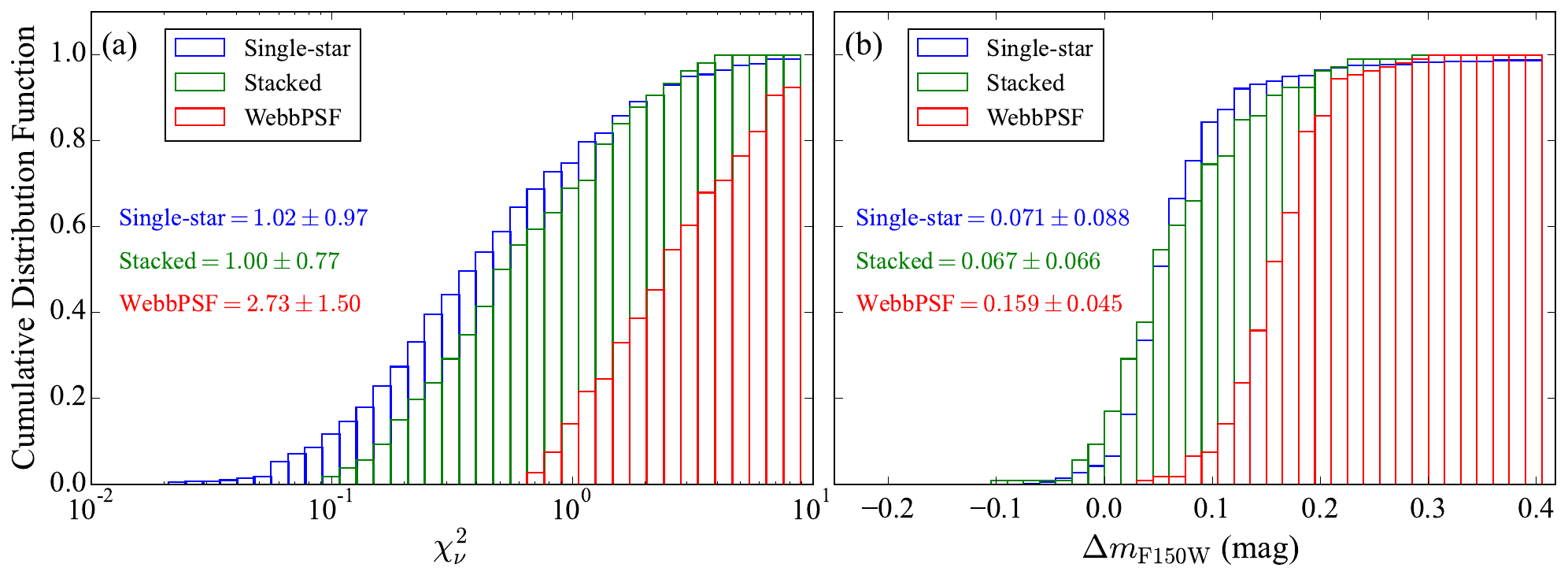}
\end{center}
\caption{The cumulative distribution function of (a) $\chi^2_{\nu}$ and (b) $\Delta m_{\rm F150W}$ for fitting stars in M\,92 observed in the F150W band (detector B1) using a PSF generated from {\tt WebbPSF} (red), a single star (blue), and stacking multiple stars (green). The median and standard deviation of the distributions of the different PSF models are given. \label{fig:psf_c}}
\end{figure*}

The NIRCam imaging data of the globular cluster M\,92 (NGC\,6341) from the JWST Resolved Stellar Populations Early Release Science program \citep{2023arXiv230104659W} provide a sufficient number of isolated, unsaturated stars to evaluate the performance of different PSF models. Although the dither pattern of the M\,92 data differs from that of CEERS, and the two data sets were not acquired simultaneously, the M\,92 data permit us to study the reliability of the PSF theoretically constructed with the {\tt Python} package {\tt WebbPSF} \citep{2014SPIE.9143E..3XP} relative to an empirical PSF generated from direct observations of stars. Moreover, we can investigate the relative performance of empirical PSFs obtained from a single star versus those produced by stacking multiple stars.

We analyze the M\,92 images observed with the F150W band in detector B1. The catalog of \cite{2022MNRAS.517..484N}\footnote{\url{https://web.oapd.inaf.it/bedin/files/PAPERs_eMATERIALs/JWST/Paper_01/}} yields 559 unsaturated, isolated stars with ${\rm AB} < 20$ mag. To account for spatial variations of the PSF, we split the detector into 25 sectors, each $409 \times 409$ pixels. Given the high density of sources in the M\,92 field, we choose a cutout size of $13 \times 13$ pixels for each star to avoid contamination from neighboring stars. For the single-star PSF models, we choose the brightest pixel of the star as the center and use it to normalize the star cutout in each sector. For the PSF models produced from stacking multiple stars, we use {\tt PSFr} \citep{2022ascl.soft10005B} to oversample the input images in each sector using third-order spline interpolation, iteratively align and co-add the images by calculating the mean pixel values, and finally resample the stacked, oversampled image back to the original pixel scale. Figure~\ref{fig:psfs} illustrates the model PSF constructed with {\tt WebbPSF}, an empirical PSF derived from an individual star, and an empirical PSF obtained from stacking multiple stars.

The star cutouts in each sector are fit using {\tt GALFITM} with the three PSF models. Figure~\ref{fig:psf_c} shows the cumulative distribution function of $\chi^2_{\nu}$ and $\Delta m_{\rm F150W}$, the difference between the best-fit model magnitude and the total integrated source flux. Over 90\% of the fits using the single-star and stacked PSFs have $\chi^2_{\nu} < 1.5$, markedly higher than the fits done using the theoretical PSF from {\tt WebbPSF}, for which only fewer than $40\%$ have $\chi^2_{\nu} < 1.5$. The empirical PSFs based on observations of real stars clearly perform significantly better than the theoretical PSF. While both the single-star and stacked PSFs show similar performance in terms of their cumulative distributions and mean values of $\chi^2_{\nu}$ and $\Delta m_{\rm F150W}$, the results using the single-star PSF exhibit larger scatter. This phenomenon is likely caused by the lower signal-to-noise ratio of the single-star PSF compared to that of the stacked PSF, rendering its performance less stable. In summary, for the analysis of the CEERS data we choose to construct our PSF model by stacking isolated, unsaturated stars in the field because they best mimic real star images and have stable performance.

\begin{figure*}[t]
\figurenum{B1}
\begin{center}
\includegraphics[width=1.0\textwidth]{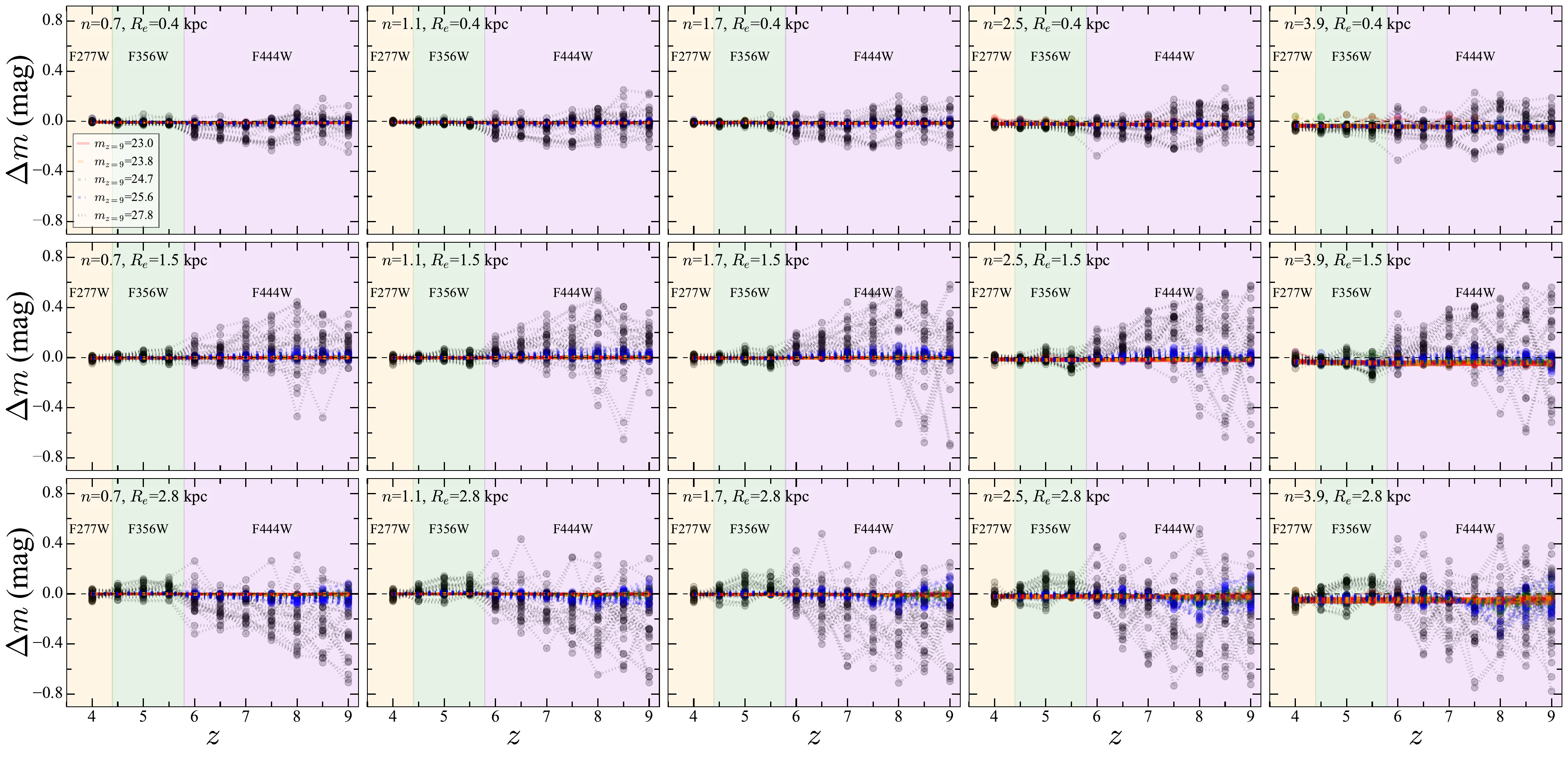}
\end{center}
\caption{The difference between measured and input values of magnitude ($\Delta m = m_{\rm out} - m_{\rm in}$) for representative simulated galaxies of different apparent magnitudes at $z = 9$, as listed in the legend of the top-left panel. The input values of $n$ and $R_e$ for each mock galaxy are in the upper-left corner of each subpanel. The differences between the measured and input values of $m$ are in the rest-frame optical band, which corresponds to different bands at different redshifts, as marked by the colored regions. \label{fig:magzeffect}}
\end{figure*}

\begin{figure*}[t]
\figurenum{B2}
\begin{center}
\includegraphics[width=1.0\textwidth]{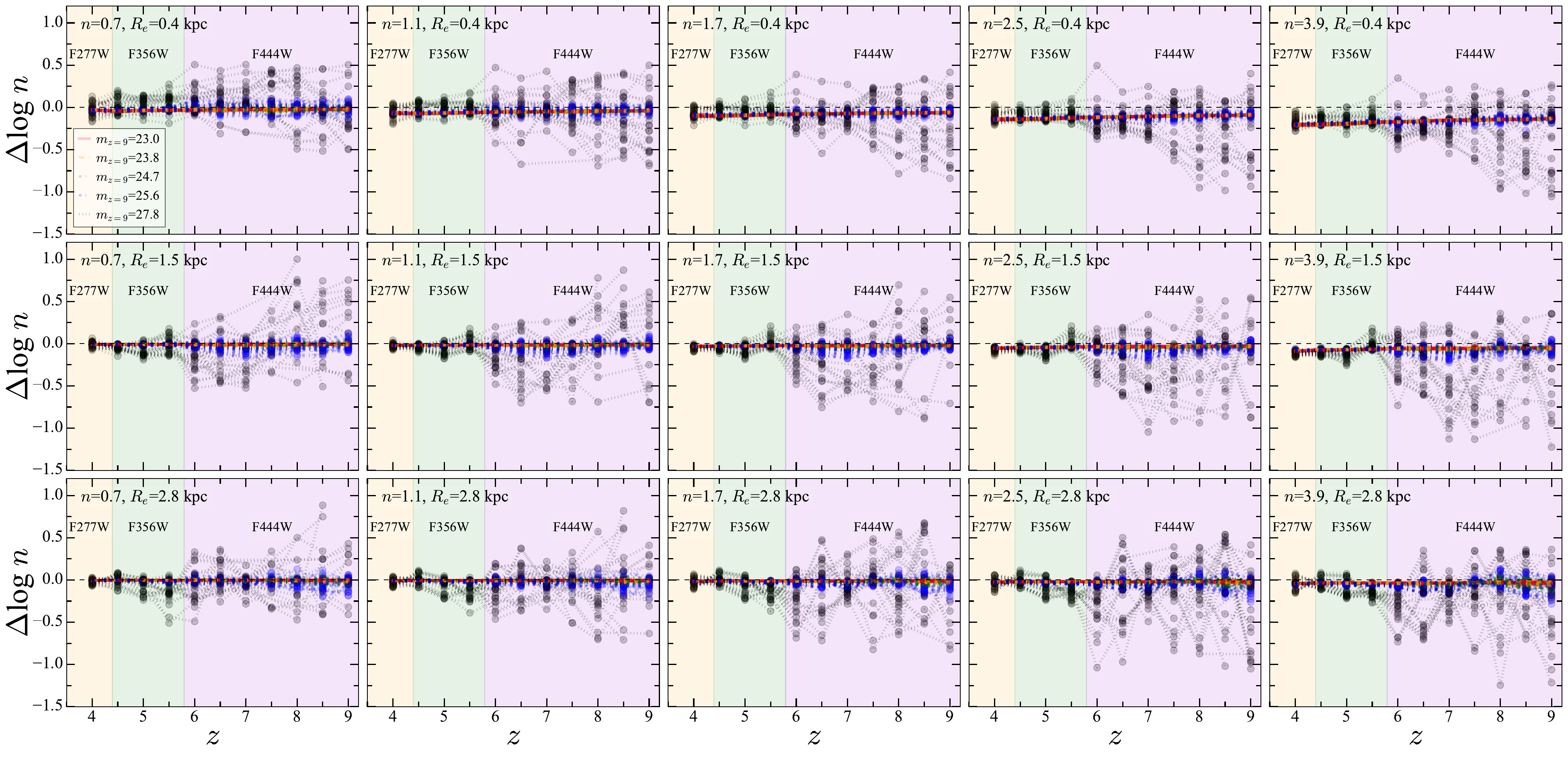}
\end{center}
\caption{As in Figure~\ref{fig:magzeffect}, but for the difference between measured and input values of S\'ersic index ($\Delta \log n = \log n_{\rm out} - \log n_{\rm in}$).
\label{fig:nzeffect}}
\end{figure*}

\begin{figure*}[t]
\figurenum{B3}
\begin{center}
\includegraphics[width=1.0\textwidth]{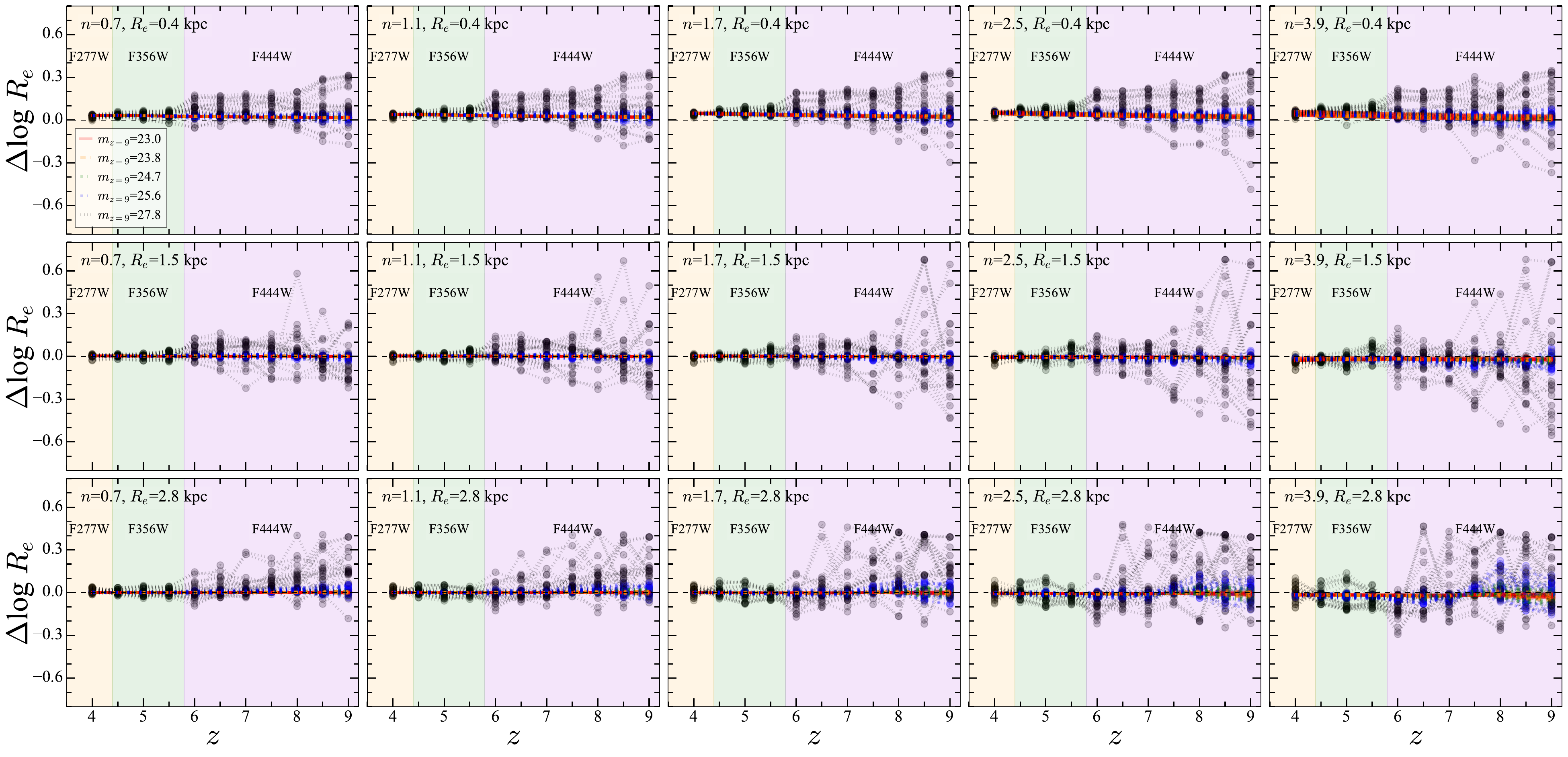}
\end{center}
\caption{As in Figure~\ref{fig:magzeffect}, but for the difference between measured and input values of effective radius ($\Delta \log R_e = \log R_{e, {\rm out}} - \log R_{e, {\rm in}}$).
\label{fig:rezeffect}}
\end{figure*}

\section{Influence of Redshift} \label{appen:zeffect}

As detailed in Section~\ref{subsec:z}, we generate images simulating the same galaxy located at different redshifts. We fit the simulated images with {\tt GALFITM} to investigate the influence of redshift on our results. Figures~\ref{fig:magzeffect}, \ref{fig:nzeffect}, and \ref{fig:rezeffect} quantify the differences between the measured and input values of magnitude $m$, S\'ersic index $n$, and effective radius $R_e$, respectively, for representative simulated galaxies at $z = 9$ with $m = 23.0$, 23.8, 24.7, 25.6, and 27.8~mag, $n = 0.7$, 1.1, 1.7, 2.5, and 3.9, and $R_e = 0.4$, 1.5, and 2.8~kpc. These input parameters span representative values observed in our sample. For simplicity, the mock images used in the analysis of redshift effects are all based on single-component S\'ersic models. We recognize that irregular features or substructures within the galaxy may lead to additional systematics not fully captured in our tests. However, this study only focuses on securing robust measurements of the integrated magnitude and the global structural parameters $n$ and $R_e$, which are not strongly influenced by the existence of substructures \citep[e.g.,][]{2010AJ....139.2097P, Meert2013, Davari2014, 2016ApJ...824..112D}. While including the effects of irregular features and substructures is beyond the scope of the current simulations, we suspect that they do not substantially alter our main conclusions.

The decrease of surface brightness with increasing redshift induces larger fluctuations at higher redshift for the differences between the recovered and input parameters. For galaxies with the same $m$ and $R_e$, the fluctuations of $\Delta m$, $\Delta \log n$, and especially $\Delta \log R_e$ become more pronounced with increasing $n$. This may be due to the difficulty of accurately modeling the central region of concentrated galaxies with large $n$. At a given $m$ and $R_e$, the fluctuations between the output and input parameter values, particularly for $n$ and $R_e$, rise toward larger $R_e$. This is because at fixed $m$ and $n$ the surface brightness of a galaxy decreases with increasing $R_e$. The deviations can be quite large at $z \approx 8-9$, especially for faint simulated galaxies with large $R_e$, whose low surface brightness renders them barely distinguishable from background fluctuations (see the black lines in the middle and bottom rows of Figures~\ref{fig:magzeffect}--\ref{fig:rezeffect}).

In general, $m$ tends to be overestimated for galaxies in our sample, possibly because we miss the faint, outer regions of these galaxies at $z>4$ even with JWST. We tend to underestimate $n$, especially for concentrated galaxies with large $n$ (right column in Figure~\ref{fig:nzeffect}). We overestimate $R_e$, particularly for small galaxies (upper row in Figure~\ref{fig:rezeffect}). These results are reasonable because cosmological surface brightness dimming impacts the inner, brighter regions of a galaxy more than its outer, fainter regions (see Equation~\ref{eq:3}), which will lead to $n$ measurements smaller and $R_e$ measurements larger than true values at high redshift. The cosmological resolution effect presents a challenge to accurate modeling of the central regions of concentrated galaxies with either large $n$ or small $R_e$. 

To determine the systematic bias and uncertainty induced by redshift effects on the photometric and structural measurements of our sample, we follow the method of \citet{Zhuang & Ho (2023)} to select simulated galaxies that are similar to our targets in the rest-frame optical band, as follows:

\begin{itemize}

\item Select mock galaxy images at the redshift closest to that of each galaxy. 

\item Select mock galaxies with brightness and structures similar to that of each galaxy, based on the error-weighted difference of $m$, $R_e$, and $n$: ${\rm \left|mock - observed\right|/\sqrt{\sigma( mock)^2+\sigma(observed)^2} < 1}$. 

\item Ensure that the axial ratios of the mock and observed galaxy do not differ by more than $q = 0.15$. 

\end{itemize}

\noindent
At least 30 mock galaxies satisfy the above criteria for each member of our sample. For each galaxy in our sample, we select the 30 simulated galaxies closest to its rest-frame optical magnitude to calculate the differences between their output and input parameter values ($\Delta m$, $\Delta \log n$, $\Delta \log R_e$). The median and standard deviation of the 30 results represents the systematic bias and uncertainty due to redshift effects, which are incorporated into the final error budget reported in Table~\ref{tab:all_par}.

\end{CJK*}
\end{document}